\documentclass[ aps, twocolumn, nofootinbib, showkeys, notitlepage, superscriptaddress]{revtex4-2}

\usepackage{natbib}
\usepackage{lipsum}
\usepackage{graphicx}
\usepackage{dcolumn}
\usepackage{bm}
\usepackage{amsmath}
\usepackage{float}
\usepackage{multirow}
\usepackage{slashed}
\usepackage{xcolor}
\usepackage{physics}
\usepackage{multirow}
\usepackage{gensymb}
\usepackage[colorlinks=true, pdfstartview=FitV, bookmarks=true, bookmarksnumbered=true, breaklinks]{hyperref}
\usepackage{mathtools}
\usepackage{lipsum}  
\usepackage{color}
\definecolor{blue}{rgb}{0.0, 0.0, 1.0}
\definecolor{red}{rgb}{1.0, 0.0, 0.0}
\definecolor{royalblue}{rgb}{0.0, 0.14, 0.4}
\hypersetup{linkcolor=blue, citecolor=blue, urlcolor=blue}

\def\orcid#1{\kern .08em\href{https://orcid.org/#1}{\includegraphics[keepaspectratio,width=0.7em]{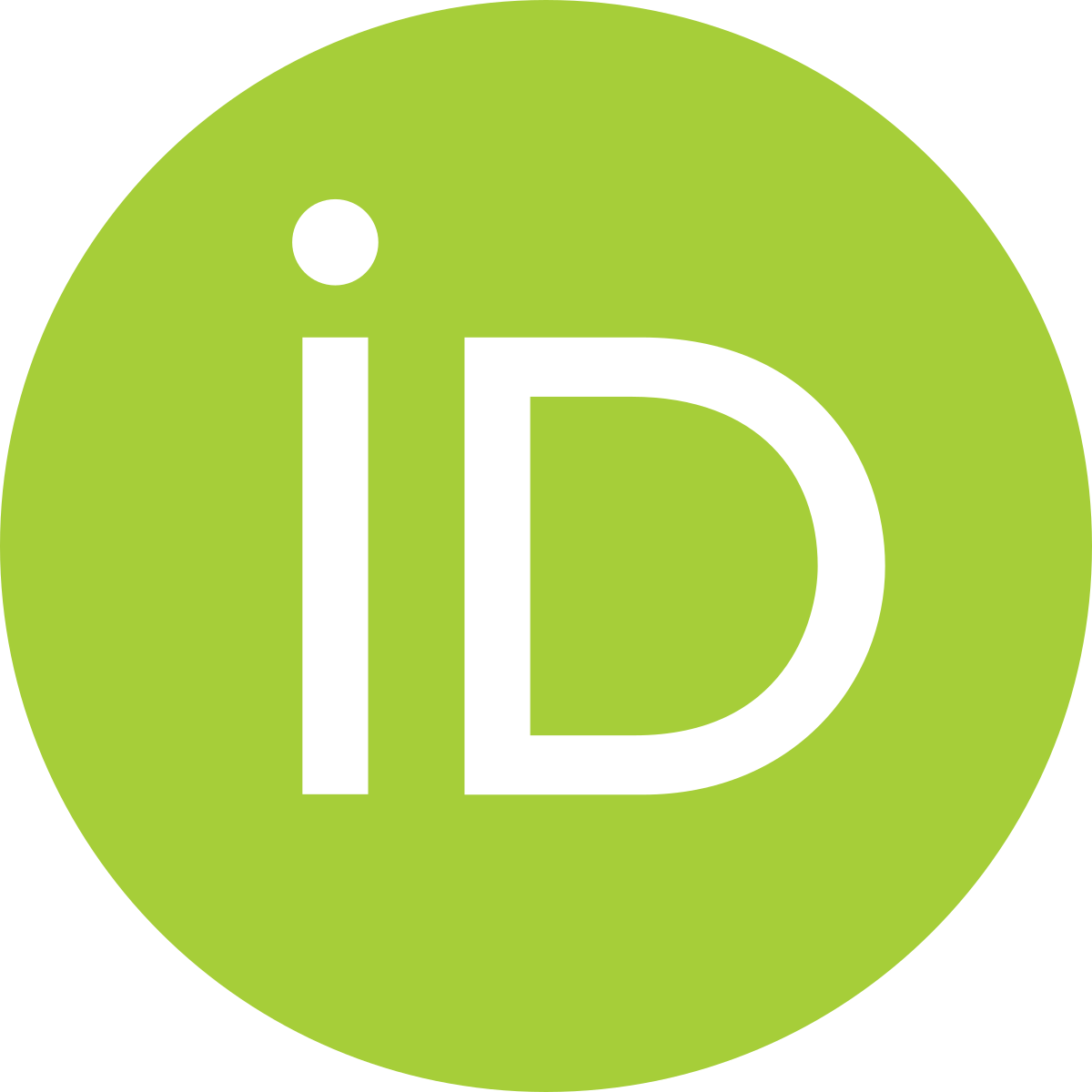}}}

\begin{document}

\title{Landau-Zener-St\"uckelberg-Majorana dynamics of magnetized quarkonia}

\author{Ahmad Jafar Arifi\orcid{0000-0002-9530-8993}}
\email{aj.arifi01@gmail.com}
\affiliation{Advanced Science Research Center, Japan Atomic Energy Agency, Tokai, Ibaraki 319-1195, Japan}
\affiliation{Research Center for Nuclear Physics, The University of Osaka, Ibaraki, Osaka 567-0047, Japan}

\author{Kei Suzuki\orcid{0000-0002-8746-4064}}
\email{k.suzuki.2010@th.phys.titech.ac.jp}
\affiliation{Advanced Science Research Center, Japan Atomic Energy Agency, Tokai, Ibaraki 319-1195, Japan}

\date{\today}

\begin{abstract}
The mass spectrum of hadrons in magnetic fields features avoided level-crossing structures arising from the mixing of spin eigenstates.
In this work, we investigate the impact of level-crossing dynamics of charmonia subjected to time-dependent magnetic fields, where we particularly focus on the occupation probabilities of two or more states as they undergo transitions at avoided crossings.
Using a static spectrum of charmonia in magnetic fields, we construct a multichannel Landau-Zener Hamiltonian.
Within this framework, we analyze the time evolution under several representative magnetic-field profiles, including linear ramps and Gaussian decays corresponding to single-passage dynamics, as well as Gaussian pulses realizing double-passage dynamics, and compute the occupation probabilities over a wide range of sweep rates and initial conditions.
Our results show that nonadiabatic dynamics, including Landau-Zener transitions and St\"uckelberg interference, strongly influences the occupation probabilities of charmonia.
These findings provide new insights into the real-time dynamics of magnetized hadrons and offer useful guidance for future lattice simulation studies.
\end{abstract}

\maketitle

\section{Introduction}

\label{sec:intro}

The behavior of hadrons under extreme external conditions, such as high temperature, density, or intense electromagnetic fields, provides a unique opportunity to deepen our understanding of the nonperturbative regime of quantum chromodynamics (QCD).
These environments are realized in nature or in experiments.
However, in some realistic situations, these conditions are time-dependent, which requires an understanding of QCD under time-dependent environments.
In particular, hadron physics in {\it static} magnetic fields has been vigorously investigated using lattice QCD simulations~\cite{Bali:2011qj,Luschevskaya:2012xd,Hidaka:2012mz,Luschevskaya:2014lga,Luschevskaya:2015cko,Bali:2017ian,Bali:2018sey,Luschevskaya:2018chr,Hattori:2019ijy,Bignell:2019vpy,Bignell:2020dze,Ding:2020hxw,Ding:2022tqn,Ding:2025pbu} (see Ref.~\cite{Endrodi:2024cqn} for a review).
On the other hand, simulations in time-dependent magnetic fields are more challenging due to the sign problem in Monte Carlo methods.
Nevertheless, in the future, sign-problem-free approaches, such as tensor-network methods~\cite{Banuls:2019rao}, will enable the simulation of real-time dynamics of hadrons. 

Heavy quarkonia, which are bound states of a heavy quark and its antiquark, serve as a simple testing ground for hadron physics.
In static magnetic fields, theoretical approaches have revealed that quarkonia undergo characteristic modifications~\cite{Marasinghe:2011bt,Tuchin:2011cg,Yang:2011cz,Tuchin:2013ie,Machado:2013rta,Alford:2013jva,Cho:2014exa,Dudal:2014jfa,Cho:2014loa,Bonati:2015dka,Sadofyev:2015hxa,Suzuki:2016kcs,Yoshida:2016xgm,Hasan:2017fmf,Singh:2017nfa, Braga:2018zlu,Iwasaki:2018pby,Amal:2018qln,Iwasaki:2018czv,Hasan:2018kvx,Braga:2019yeh,Hasan:2020iwa,Zhou:2020ssi,Braga:2020hhs,Braga:2021fey,Jena:2022nzw,Hu:2022ofv,Ghosh:2022sxi,Parui:2022msu,Sebastian:2023tlw,Nilima:2024nvd,Jena:2024cqs,Shukla:2024qlf,Wen:2025dwy,Jena:2025xcf,Arifi:2025ivt,Dominguez:2025nar} (see Refs.~\cite{Hattori:2016emy,Zhao:2020jqu,Iwasaki:2021nrz} for reviews).
One of the features is \emph{level repulsions} due to mixing between eigenstates with different spin quantum numbers.
This is induced by the coupling between a magnetic field and quark magnetic moments.
For example, pseudoscalar mesons (e.g., $\eta_c$, $\eta_c^\prime$, and $\eta_b$) are mixed with the corresponding vector partners (e.g., $J/\psi$, $\psi^\prime$, and $\Upsilon$)~\cite{Yang:2011cz,Alford:2013jva}.
As a related phenomenon, the mass spectrum in the strong-field region shows complex \emph{avoided level crossings}~\cite{Suzuki:2016kcs,Yoshida:2016xgm}.
Such a level structure is a key point in our current work.

Despite substantial progress in the static case, the dynamical response of quarkonia to time-dependent magnetic fields remains comparatively less explored~\cite{Guo:2015nsa,Suzuki:2016fof,Dutta:2017pya,Hoelck:2017dby,Bagchi:2018olp,Iwasaki:2021kms}.
These previous works were primarily motivated by realistic heavy-ion collision experiments, where the magnetic field changes rapidly. In contrast, our motivation is more general: to gain a broader understanding of the time evolution of hadrons and to design physical systems that will be valuable for future lattice QCD simulations.
In particular, we focus on that charmonia undergo \emph{nonadiabatic transitions} as the system sweeps through avoided crossings.

Nonadiabatic transitions are automatically included when a time-dependent Schr\"odinger equation with coupled channels is solved, as in the previous works~\cite{Guo:2015nsa,Dutta:2017pya,Bagchi:2018olp,Iwasaki:2021kms}.
In this work, we develop a new approach based on a \emph{multichannel Landau-Zener (LZ) model}.
In general, the nonadiabatic transition around an avoided crossing can be described as the well-known \emph{LZ  transition}~\cite{Majorana:1932,Landau:1932wdt,Landau:1932vnv,Zener:1932,Stuckelberg:1932}\footnote{The Landau-Zener transition represents a single-passage problem, while two-passage problems exhibit the St\"uckelberg interference.
To recognize the contributions of the four researchers who studied this problem in 1932, such phenomena are collectively referred to as Landau-Zener-St\"uckelberg-Majorana (LZSM) transitions~\cite{Ivakhnenko:2022sfl}.} (see Ref.~\cite{Shevchenko:2010ms,Ivakhnenko:2022sfl} for reviews), where a transition is characterized by only two parameters, (i) the \emph{sweep rate} of a time-dependent external parameter and (ii) the \emph{gap parameter} at the avoided crossing. 
Note that magnetized quarkonia exhibit an avoided-crossing gap on the order of tens of MeV, which is much larger than other LZ systems in atomic and condensed matter physics.
Our key idea is to construct an effective time-dependent LZ-type Hamiltonian using the magnetic-field dependence of a charmonium spectrum.
Once this effective Hamiltonian is constructed, we can solve the time-dependent Schr\"odinger equation under specific magnetic-field profiles. 

Our approach has several key advantages:
(i) The analysis of a two-channel model provides an intuitive picture using well-known analytic solutions. This allows us to understand the transitions with only two types of parameters, which is a significant advantage compared to the time evolution of the original Hamiltonian~\cite{Iwasaki:2021kms}.
(ii) The model enables us to extract the LZ parameters directly from the charmonium spectrum alone, without requiring any knowledge of the underlying full Hamiltonian. (iii) It also provides a natural setting to compare the \emph{diabatic} and \emph{adiabatic} bases, allowing us to clearly evaluate the degree of adiabaticity.

The remainder of this paper is organized as follows. Section~\ref{sec:model} reviews the calculation of the quark-model spectra in a static magnetic field, followed by the formulation of the time-dependent multichannel LZ Hamiltonian and the extraction of its parameters from the static spectra.
Section~\ref{sec:applications} provides numerical results for representative magnetic-field profiles and discusses their physical implications. Section~\ref{sec:conclusion} summarizes our findings and outlooks.
In Appendix~\ref{sec:analytic}, we compare the known analytic solutions with our numerical simulations to verify the correctness of our results.

\section{Model description}
\label{sec:model}

In this section, we briefly review the calculation of charmonium spectra in a quark model in static magnetic fields~\cite{Suzuki:2016kcs,Yoshida:2016xgm}. 
Then, we construct the multichannel LZ Hamiltonian and determine its parameters by fitting to the static spectra, which allows us to study the dynamical evolution of quarkonia in time-dependent magnetic fields. 

\subsection{Charmonia in static magnetic fields}

We study charmonium states within a nonrelativistic constituent quark model under the influence of an external magnetic field $\bm{B} = B \hat{z}$. 
The two-body Hamiltonian for a charm-anticharm system is given by~\cite{Machado:2013rta,Alford:2013jva}
\begin{equation}
    H_{c\bar{c}} = \sum_{i=c,\bar{c}} \left[ m_i + \frac{(\bm{p}_i - q_i \bm{A})^2}{2m_i} - \bm{\mu}_i \cdot \bm{B} \right] + V_{c\bar{c}}(r),
\end{equation}
where $m_i$ and $\bm{p}_i$ are the mass and momentum of the $i$ th (anti)charm quark, $q_i$ is its electric charge, $\bm{A}$ is the vector potential of the magnetic field, and $\bm{\mu}_i$ denotes the quark magnetic moment.  
We adopt the symmetric gauge, $\bm{A}(\bm{r}_i) = \frac{1}{2}\bm{B}\times \bm{r}_i$, and take the quark magnetic moment as $\bm{\mu}_i = g q_i \bm{S}_i / 2 m_i$ with Lande $g$-factor $g=2$ and the quark spin vector $\bm{S}_i$.

The quark-antiquark potential $V_{c\bar{c}}$ is modeled as~\cite{Eichten:1974af,Barnes:2005pb}  
\begin{equation}
V_{c\bar{c}}(r) = C + \sigma r - \frac{A}{r} + \beta (\bm{S}_c \cdot \bm{S}_{\bar{c}}) e^{-\Lambda r^2},
\end{equation}
where the relative coordinate is $r = \sqrt{z^2 + r_\perp^2}$, $\sigma$, $A$, and $\beta$ characterize the confining, Coulomb, and spin-spin potentials, respectively, and $\Lambda$ parametrizes the Gaussian smearing~\cite{Barnes:2005pb} of the spin-spin term.  
$C$ is an energy shift adjusting the overall spectrum.
The spin operator expectation values yield $1/4$ and $-3/4$ for vector (spin-triplet) and pseudoscalar (spin-singlet) mesons, respectively.
Here we adopt the same parameters in the previous works~\cite{Suzuki:2016kcs,Yoshida:2016xgm}, $m_c=1.784$ GeV,
$A=0.713$, $\sqrt{\sigma} = 0.402$ GeV, $\beta =0.4778$ GeV, $\Lambda= 1.020$ GeV$^2$, and $C=-0.5693$ GeV.

For a charmonium ($q_1=-q_2=q$) with vanishing total pseudomomentum $\bm{K}=0$, the Hamiltonian reduces to the relative-motion form~\cite{Machado:2013rta,Alford:2013jva,Andreichikov:2013zba}
\begin{equation}\label{eq:H_rel}
    H = -\frac{\bm{\nabla}^2}{2\mu} + \frac{q^2 B^2}{8\mu} r_\perp^2 + V(r) - \sum_{i=q,\bar{q}} \bm{\mu}_i \cdot \bm{B},
\end{equation}
where $\mu = m_q/2$ is the reduced quark mass.  
The second term represents the transverse potential arising from the quark-Landau-level contribution.
This Hamiltonian preserves the cylindrical symmetry in the $x$-$y$ plane and reflection symmetry along $z$.

The magnetic interaction mixes the longitudinal spin component of the triplet state, $\ket{S=1, S_z=0} = \ket{10}$, with the singlet state, $\ket{S=0, S_z=0} = \ket{00}$, while the transverse components $\ket{S=1, S_z=\pm1}$ remain decoupled. 
The off-diagonal matrix element is
\begin{equation}
\bra{10} - (\bm{\mu}_1 + \bm{\mu}_2) \cdot \bm{B} \ket{00} = - \frac{g q B}{4 \mu}.
\end{equation}
This off-diagonal mixing leads to avoided level crossings between the pseudoscalar and longitudinal vector states as the magnetic field strength is varied. 

Finally, the Hamiltonian in Eq.~\eqref{eq:H_rel} leads to a coupled-channel Schr\"odinger equation for the $S_z=0$ components. 
This equation can be solved numerically using the \emph{cylindrical Gaussian expansion method} (CGEM)~\cite{Suzuki:2016kcs,Yoshida:2016xgm}, which is an extension of the standard Gaussian expansion method~\cite{Kamimura:1988zz,Hiyama:2003cu} adapted to systems with cylindrical symmetry. 
The CGEM allows us to obtain the charmonium spectrum in static magnetic fields with high accuracy.

\subsection{Multichannel Landau-Zener Hamiltonian}

To describe the time evolution of the system, we begin with the time-dependent Schr\"odinger equation
\begin{equation}\label{eq:tim-dep}
i\,\frac{d}{dt}\ket{\psi(t)} = H(t)\ket{\psi(t)},
\end{equation}
where $\ket{\psi(t)}$ is an $N$-component state vector representing all relevant charmonium channels. 
Once $H(t)$ is specified, this equation allows us to evaluate the time-dependent \emph{occupation probabilities} among different states under any chosen magnetic-field profile.

Although one could in principle construct $H(t)$ directly from the full quark-model Hamiltonian, including the explicit time dependence induced by the magnetic field as practiced in the previous study~\cite{Iwasaki:2021kms}. 
In this work, we propose a different approach. 
To extend our study to time-dependent magnetic fields while keeping the dynamics tractable, we map the static charmonium spectrum of the quark model onto a multichannel LZ Hamiltonian.

In this effective description, the magnetic-field dependence appears only in the diagonal terms, while the off-diagonal couplings, responsible for mixing at avoided crossings, are treated as constants. 
Although approximate, this construction is particularly well suited for studying nonadiabatic transitions. 
The parameters of the LZ Hamiltonian can be fixed directly from the quark-model spectrum, and the formalism naturally exposes the relationship between diabatic and adiabatic bases, making the physics of level crossings and transition probabilities especially transparent.

\begin{figure*}[t]
    \centering
    \includegraphics[width=0.47\textwidth]{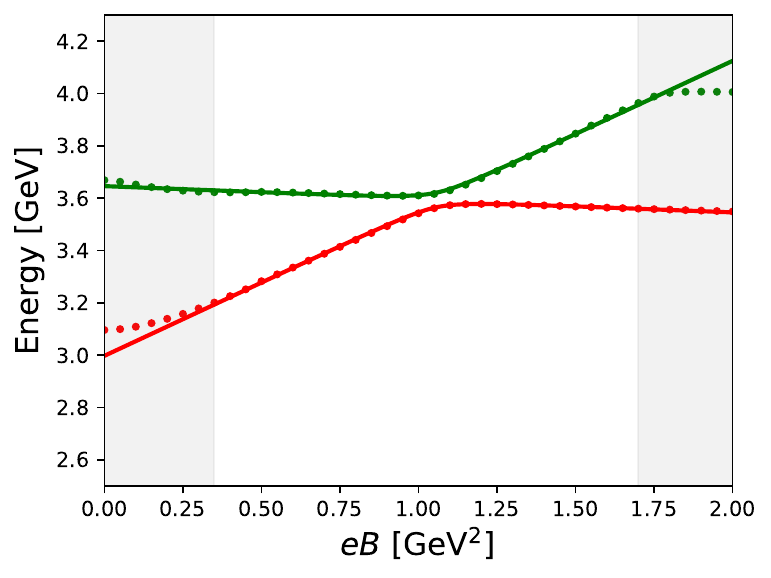}
    \includegraphics[width=0.47\textwidth]{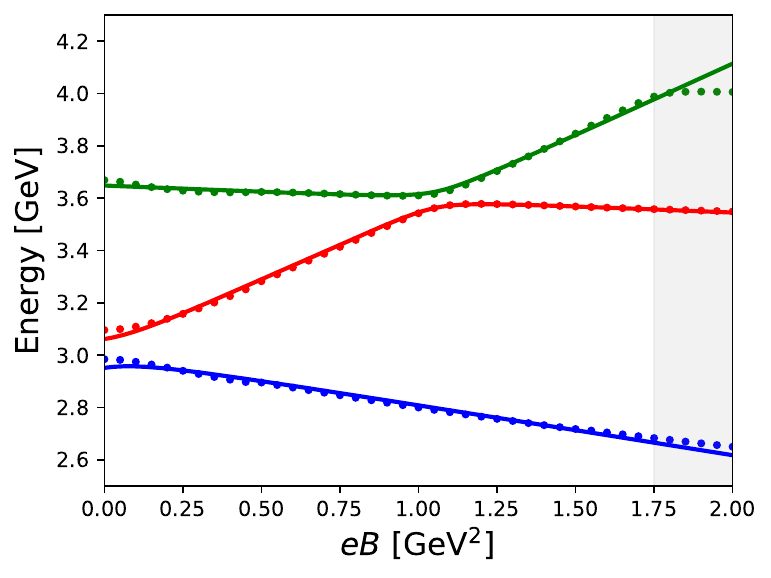}
    \includegraphics[width=0.47\textwidth]{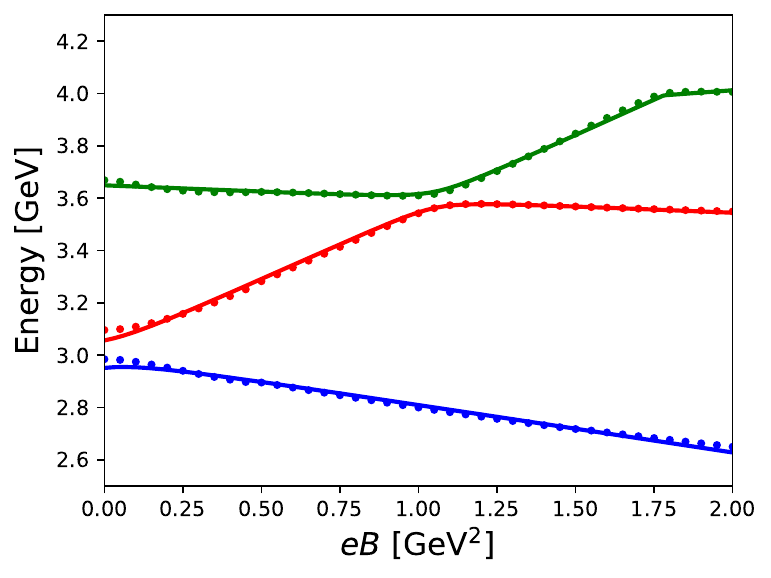}
    \includegraphics[width=0.47\textwidth]{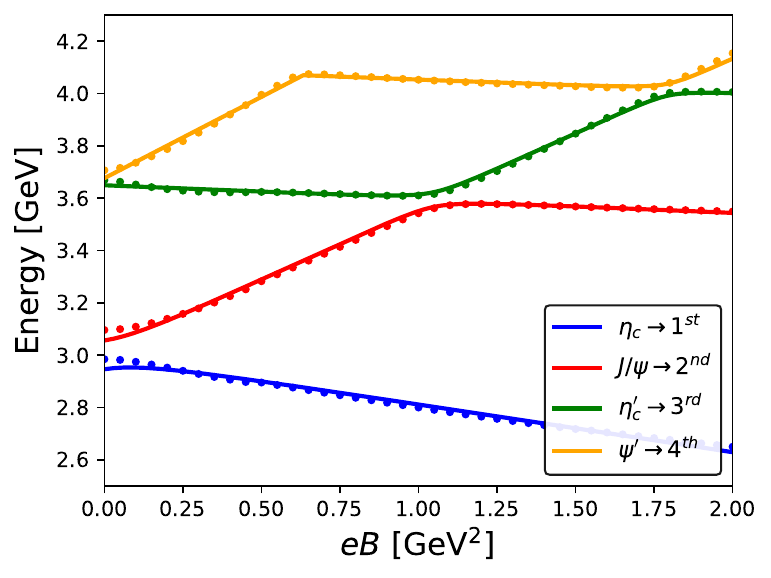}
\caption{Comparison between the eigenvalues of the multichannel LZ Hamiltonians and the quark-model spectrum for the two-, three-, four-, and five-channel cases.
Each panel shows the fitted LZ energy levels (solid lines) together with the discrete quark-model data points (markers), demonstrating good agreement in the region of interest.
The fitting parameters, energy slopes $\alpha_i$, offsets $\delta_i$, and off-diagonal couplings $\Delta_{i,i+1}$, are listed in Table~\ref{tab:LZ_parameters_full}.
The gray regions indicate data excluded from the fit.
Including more channels enables the LZ model to reproduce additional avoided crossings and higher excited states. 
The highest state in the four- and five-channel cases is omitted because it does not produce the correct spectrum. To reproduce it, one additional channel is required.
 }
    \label{fig:fitting}
\end{figure*}

\renewcommand{\arraystretch}{1.5}  

\begin{table*}[t!] 
\begin{ruledtabular}
\caption{Parameters of the multichannel LZ Hamiltonians for the two–five channel models.
The slopes $\alpha_i$, level offsets $\delta_i$, and nearest-neighbor couplings $\Delta_{i,i+1}$ are fitted to reproduce the charmonium spectrum.
The $\alpha_i$ parameters are given in GeV$^{-1}$, while $\delta_i$ and $\Delta_{i,i+1}$ are expressed in GeV. 
The two-channel parameters differ from the others because this model starts from the $J/\psi$ rather than the $\eta_c$ state.
} 
\label{tab:LZ_parameters_full}
\begin{tabular}{c|ccccc|ccccc|cccc}
Model & $\alpha_1$ & $\alpha_2$ & $\alpha_3$ & $\alpha_4$ & $\alpha_5$ &  $\delta_1$&  $\delta_2$ & $ \delta_3$ & $ \delta_4$ & $ \delta_5$ &
$\Delta_{1,2}$ & $\Delta_{2,3}$ & $\Delta_{3,4}$ & $\Delta_{4,5}$ \\ \hline

2-channel 
 & 0.562 & $-$0.049 & $\dots$ & $\dots$ & $\dots$
 & 2.999 & 3.645 & $\dots$ & $\dots$ & $\dots$
 & 0.028 & $\dots$ & $\dots$ & $\dots$ \\

3-channel 
 & $-$0.192 & 0.550 & $-$0.050 & $\dots$ & $\dots$
 & 3.005 & 3.010 & 3.647 & $\dots$ & $\dots$
 & 0.055 & 0.031 & $\dots$ & $\dots$ \\

4-channel 
 & $-$0.183 & 0.548 & $-$0.051 & 0.088 & $\dots$
 & 2.996 & 3.013 & 3.649 & 3.836 & $\dots$
 & 0.052 & 0.031 & $-$0.012 & $\dots$ \\

5-channel 
 & $-$0.177 & 0.553 & $-$0.065 & 0.0017 & 0.626
 & 2.991 & 3.011 & 3.798 & 3.905 & 3.677
 & 0.053 & 0.041 & 0.206 & 0.006 \\

\end{tabular}
\end{ruledtabular}
\renewcommand{\arraystretch}{1}  
\end{table*}

To implement this idea, we parametrize the Hamiltonian in the following form:
\begin{widetext}
\begin{equation}
H(t) =
\begin{pmatrix}
\alpha_1 eB(t) + \delta_1 & \Delta_{1,2} & 0 & \cdots & 0 \\
\Delta_{1,2} & \alpha_2 eB(t) + \delta_2 & \Delta_{2,3} & \cdots & 0 \\
0 & \Delta_{2,3} & \alpha_3 eB(t) + \delta_3 & \cdots & 0 \\
\vdots & \vdots & \vdots & \ddots & \Delta_{N-1,N} \\
0 & 0 & 0 & \Delta_{N-1,N} & \alpha_N eB(t) + \delta_N
\end{pmatrix},
\end{equation}
\end{widetext}
where $N$ denotes the number of coupled channels. 

The diagonal elements are now parametrized as
\begin{equation}
H_{ii}(t) = \alpha_i eB(t) + \delta_i,
\end{equation}
where $\alpha_i$ represents the slope of the energy level with respect to the magnetic field, $B(t)$ is the time-dependent external magnetic field profile,
and $\delta_i$ is the state-dependent offset.
The off-diagonal elements $\Delta_{i,i+1}$ are the couplings between neighboring charmonium states.  
These couplings are assumed to be constant.  
For example, in the two-channel case, the Hamiltonian reduces to
\begin{equation}
H(t) =
\begin{pmatrix}
\alpha_1 eB(t) + \delta_1 & \Delta_{1,2} \\
\Delta_{1,2} & \alpha_2 eB(t) + \delta_2
\end{pmatrix},
\end{equation}
while three- to five-channel generalizations incorporate additional states and couplings.

Within this parametrization, the \emph{LZ sweep rate} arises naturally from the time dependence of the diabatic level spacing. It is defined as the time derivative of the level spacing, $\nu_\mathrm{LZ} = d(E_1 - E_2)/dt$. When the energies depend on an external magnetic field $B(t)$, and the field varies linearly as $B(t)=\nu_B t$, the sweep rate becomes $\nu_\mathrm{LZ} = |\alpha_1 - \alpha_2|\nu_B $, where $\alpha_i$ are the slopes of the diabatic energy levels with respect to $B$ and $\nu_B$ is the magnetic field sweep rate.

\subsection{Fitting of the Landau-Zener Hamiltonians}
\label{sec:fitting}

To connect the static quark-model spectrum with the time-dependent LZ model, we fit the discrete charmonium energy levels obtained in our previous works~\cite{Suzuki:2016kcs,Yoshida:2016xgm} using multichannel LZ Hamiltonians.
Note that the quark-model results serve as an example in our approach, while input from other models or lattice simulations can also be used in the future.
Therefore, our analysis demonstrates how a static spectrum can be systematically mapped onto the LZ Hamiltonian.

Figure~\ref{fig:fitting} compares the eigenvalues of the fitted LZ Hamiltonians with the corresponding quark-model energies for the two-, three-, four-, and five-channel cases.
Note that the gray regions are excluded from the fit for the two- and three-channel models.
In the present analysis, the coupled-channel model space is built from the charmonium states
$\{\eta_c,\, J/\psi,\, \eta_c^\prime,\, \psi^\prime,\, \eta_c^{\prime\prime}\}$.
Specifically, $N=2$ includes $\{J/\psi, \eta_c^\prime\}$; $N=3$ further includes $\eta_c$;
$N=4$ adds $\psi^\prime$; and $N=5$ includes $\eta_c^{\prime\prime}$.

The parameters extracted from this fitting procedure are summarized in Table~\ref{tab:LZ_parameters_full}.
We also note that the parameter values for the two-channel model differ from those of the higher-channel models.
This is because now the two-channel model starts with the $J/\psi$-$\eta_c^\prime$, not with the $\eta_c$-$J/\psi$.

The two-channel model captures the lowest avoided crossing between the $J/\psi$ and $\eta_c^\prime$-started states at $eB \sim 1$ GeV$^2$.
The slope parameters, $\alpha_1 = 0.562$ GeV$^{-1}$ and $\alpha_2 = -0.049$ GeV$^{-1}$, determine how rapidly each energy level changes with the magnetic field $eB$, while the coupling $\Delta_{1,2} = 0.028$ GeV sets the size of the avoided crossing gap.
The level offsets, $\delta_1 = 2.999$ GeV and $\delta_2 = 3.645$ GeV, shift the positions of the crossing along the energy axis.
We find that this minimal model  accurately reproduces the lowest avoided crossing but cannot describe the spectrum far from that.

\begin{figure*}[t!]
    \centering
    \includegraphics[width=0.96\textwidth]{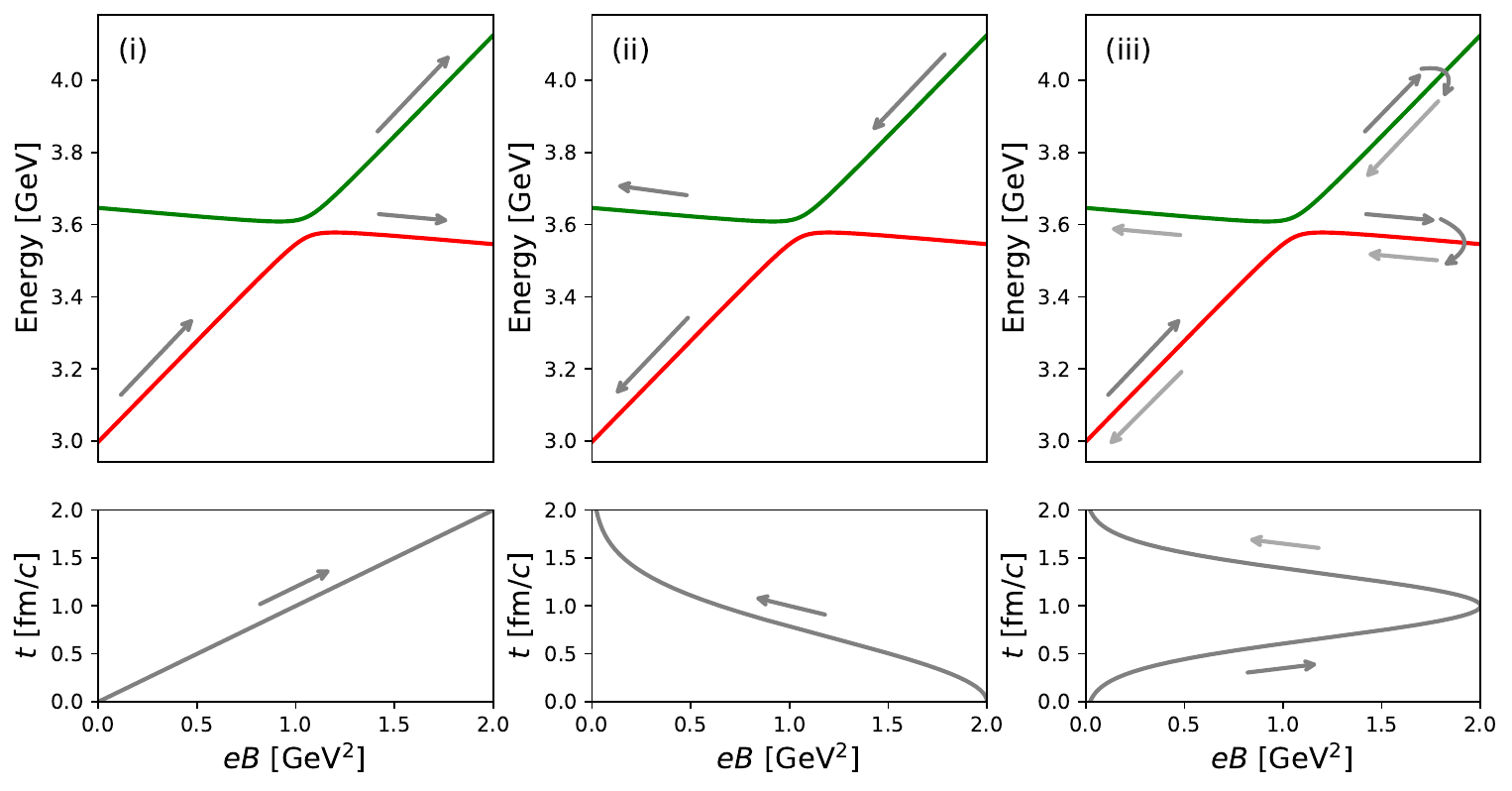}
    \caption{Schematic illustration of the avoided crossings in the two-channel model and the driving magnetic-field profiles considered in this work: (i) linear ramps, (ii) Gaussian decays, and (iii) Gaussian pulses. Panels (i) and (ii) correspond to single passages through the avoided crossing, while panel (iii) shows a double passage. The arrows indicate the direction of time evolution along the magnetic-field trajectories. }
    \label{fig:schematic}
\end{figure*}

Adding a third channel incorporates the ground state starting from $\eta_c$.
In this three-channel model, the coupling $\Delta_{2,3} = 0.031$ GeV is analogous to $\Delta_{1,2}$ in the two-channel model.
Thus, the magnitude of gap is a robust parameter characterizing the avoided crossing, although it is slightly depends on the number of channels.
Furthermore, the gap controlled by $\Delta_{1,2}$ in the three-channel model allows it to reproduce the level repulsion between $\eta_c$ and $J/\psi$ at weak magnetic fields.
Adding further channels in the four- and five-channel models introduces additional slopes, offsets, and couplings (see Table~\ref{tab:LZ_parameters_full}) to capture higher excited states and additional avoided crossings, systematically improving the agreement with the quark-model spectrum. It is important to note that, to correctly reproduce all four low-lying states, five channels are required.

Therefore, the fitting procedure demonstrates that the LZ Hamiltonian can effectively capture the quark model spectrum by adjusting a relatively small set of physically motivated parameters. 
This provides a solid basis for analyzing the dynamical evolution of charmonium in time-dependent magnetic fields. 
In the following, we focus on applications of the two- and three-channel Hamiltonians, which allow a detailed discussion of the transition dynamics. 
For the four- and five-channel cases, we provide brief remarks, as the occupation probabilities become increasingly complex and harder to interpret visually.

\section{Applications}
\label{sec:applications}

In this section, we discuss the application of the multichannel LZ Hamiltonian for studying dynamical transitions among charmonium states.
We consider three representative time-dependent magnetic-field profiles, as illustrated schematically in Fig.~\ref{fig:schematic}: a linear ramp, a Gaussian decay, and a Gaussian pulse. 
The linear ramp and Gaussian decay correspond to a \emph{single passage} through an avoided crossing, while the Gaussian pulse represents a \emph{double passage}. 
The arrows in the figure indicate the direction of time evolution along the magnetic-field trajectories.

The time dependence of the external magnetic field is modeled by the explicit function $eB(t)$, which serves as the control parameter governing the evolution of the multichannel Hamiltonian. 
To explore the sensitivity of physical observables to the field dynamics, the three profiles are defined as
\begin{align}
    eB(t) =
    \begin{cases}
       \nu_B t,  & \text{Linear ramp}, \\[4pt]
       eB_\mathrm{max}
       \exp\!\left[-\dfrac{t^2}{2\gamma_D^2}\right],  & \text{Gaussian decay}, \\[4pt]
       eB_\mathrm{max}
       \exp\!\left[-\dfrac{(t - t_\mathrm{peak})^2}{2\gamma_P^2}\right],  & \text{Gaussian pulse}.
    \end{cases}
\end{align}
Here, $\nu_B=eB_\mathrm{max}/t_\mathrm{max}$ sets the slope of the linear ramp, $B_\mathrm{max}$ denotes the maximum field strength, $\gamma_D$ and $\gamma_P$ control the temporal widths of the decaying and pulse profiles, respectively, and $t_\mathrm{peak}$ specifies the time at which the pulse reaches its maximum.
All profiles are defined for $0 \le t \le t_\mathrm{max}$, covering the full evolution.

\begin{figure*}[t!]
    \centering
    \includegraphics[width=0.96\textwidth]{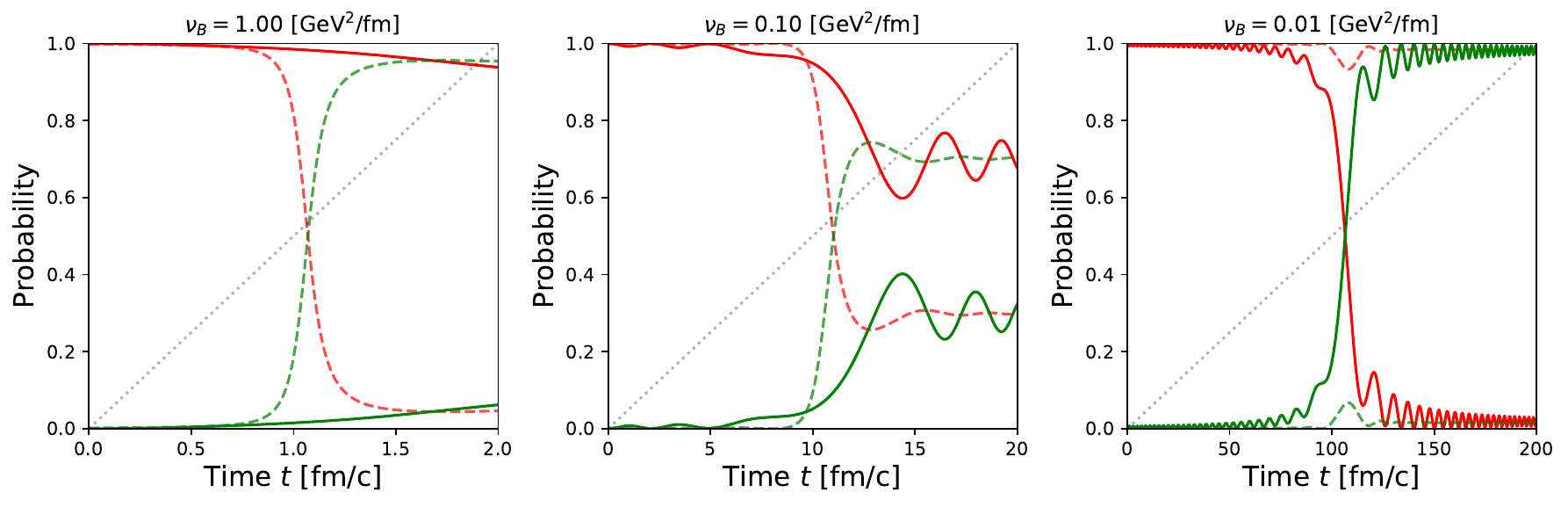}
    \includegraphics[width=0.96\textwidth]{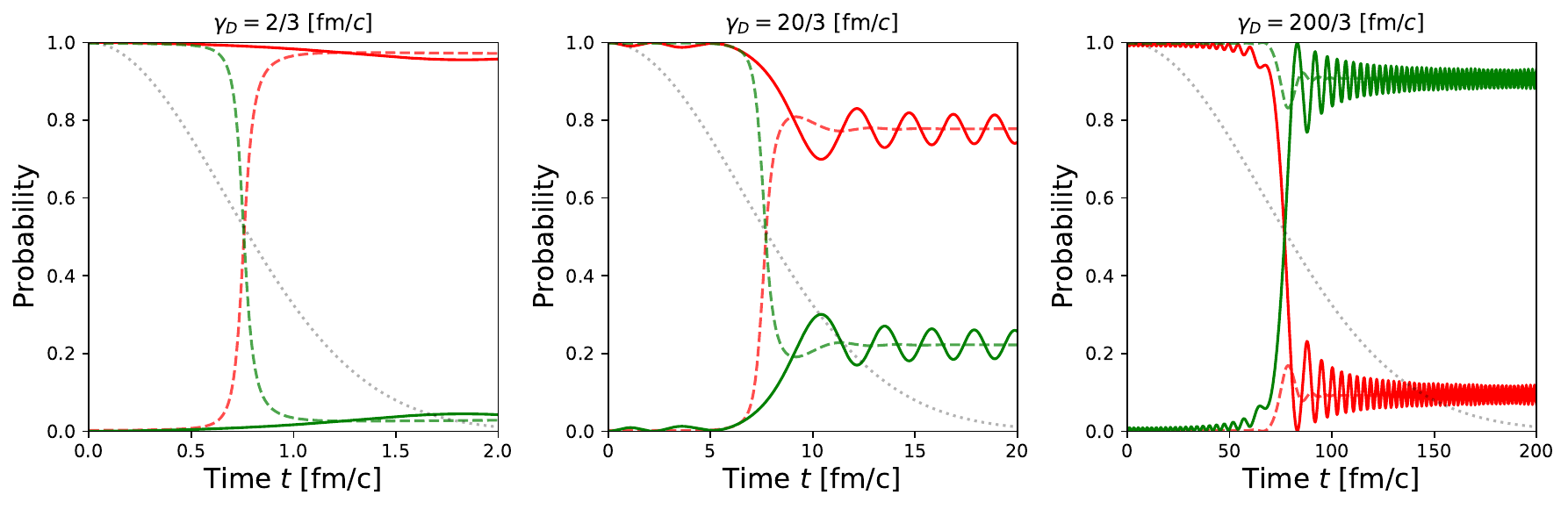}
\caption{Occupation probabilities in the two-channel model for a single passage at three sweep rates with a $\psi=(1,0)^T$ initial state: (top panels) linear ramp and (bottom panels) Gaussian decay.
The gray dotted line shows the magnetic-field profiles with a maximum value of $eB_{\max}=2~\text{GeV}^2$. 
The red and green lines represent the occupation probabilities of the $J/\psi$- and $\eta_c(2S)$-started states, respectively, shown in the diabatic basis (solid lines) and in the adiabatic basis (dashed lines). In general, the line style and color codes for the occupation probabilities follow the notation in Table~\ref{tab:state-shorthand}.
}
    \label{fig:single-2level}
\end{figure*}

\subsection{Occupation probabilities}

Before proceeding further, it is crucial to explain the different definition of the occupation probabilities.
They are analyzed in both the diabatic basis, corresponding to the uncoupled Hamiltonian, and the adiabatic basis, corresponding to the instantaneous eigenstates of the coupled system.
Magnetic-field-induced mixing produces an avoided crossing in the adiabatic energies, and whether the system follows diabatic or adiabatic paths is determined by the effective sweep rate through this region. 
To simplify the discussion, we introduce the shorthand notation for the occupation probabilities given in Table.~\ref{tab:state-shorthand}.

\begin{table}[t!]
\centering
\caption{Shorthand notation for adiabatic and diabatic probabilities of charmonium states, with line style and color coding as used in the plots.}
\begin{tabular}{c c c c c}
\hline\hline
State & Basis & Line Style & Color & Shorthand \\
\hline
$\eta_c$-started & Diabatic & Solid & Blue & $P_{\rm dia}[{\eta_c}]$ \\
     & Adiabatic & Dashed &  & $P_{\rm adi}[{\eta_c}]$ \\
$J/\psi$-started & Diabatic & Solid & Red & $P_{\rm dia}[{J/\psi}]$ \\
& Adiabatic & Dashed & & $P_{\rm adi}[{J/\psi}]$ \\
$\eta_c^\prime$-started & Diabatic & Solid & Green & $P_{\rm dia}[{\eta_c^\prime}]$ \\
& Adiabatic & Dashed &  & $P_{\rm adi}[\eta_c^\prime]$ \\
$\psi^\prime$-started & Diabatic & Solid & Orange & $P_{\rm dia}[{\psi^\prime}]$ \\
& Adiabatic & Dashed &  & $P_{\rm adi}[\psi^\prime]$ \\
\hline\hline
\end{tabular}
\label{tab:state-shorthand}
\end{table}

In practice, the occupation probabilities in the diabatic basis are directly obtained from the components of the time-evolved state vector,
\begin{equation}
P_{\rm dia}^{(i)}(t) = |\psi_i(t)|^2,
\end{equation}
where $\psi_i(t)$ is the $i$th component of the solution to the time-dependent Schrödinger equation in Eq.~\eqref{eq:tim-dep}.
By contrast, the probabilities in the adiabatic basis are computed by projecting the time-evolved state onto the instantaneous eigenstates of the Hamiltonian,
\begin{equation}
P_{\rm adi}^{(i)}(t) = \left|\braket{\phi_{\rm adi}^{(i)}(t)}{\psi(t)}\right|^2,
\end{equation}
where $\phi_{\rm adi}^{(i)}(t)$ denotes the $i$th instantaneous eigenstate at time $t$.

\begin{figure*}[t!]
    \centering
    \includegraphics[width=0.96\textwidth]{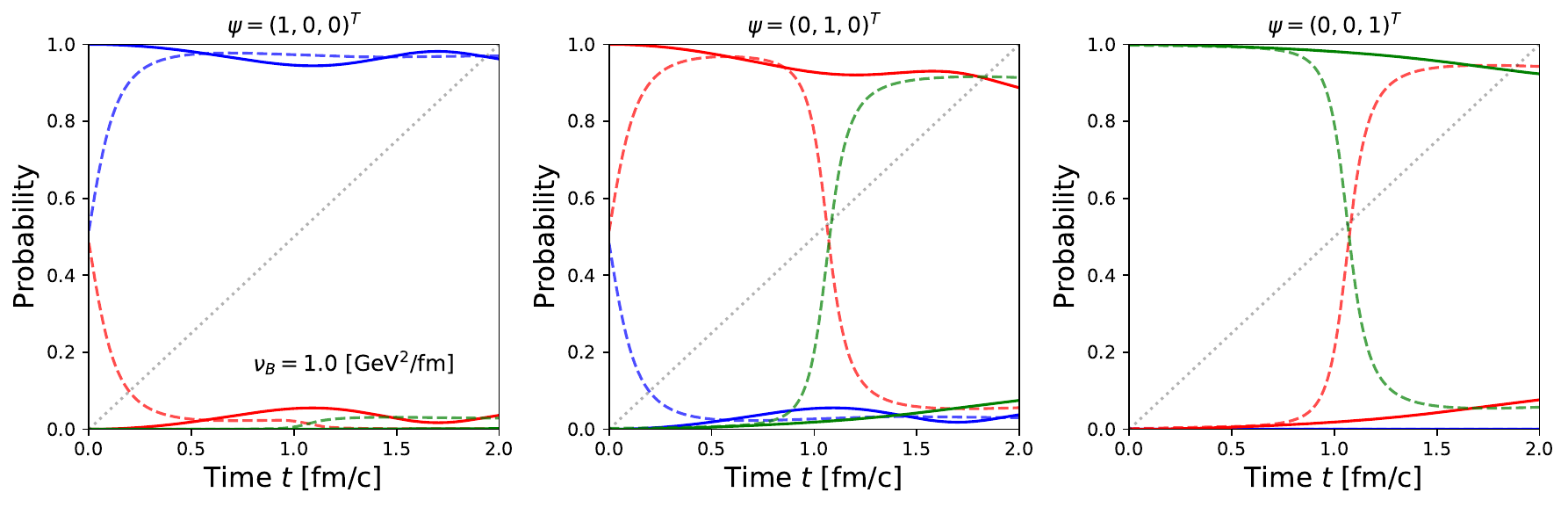}
    \includegraphics[width=0.96\textwidth]{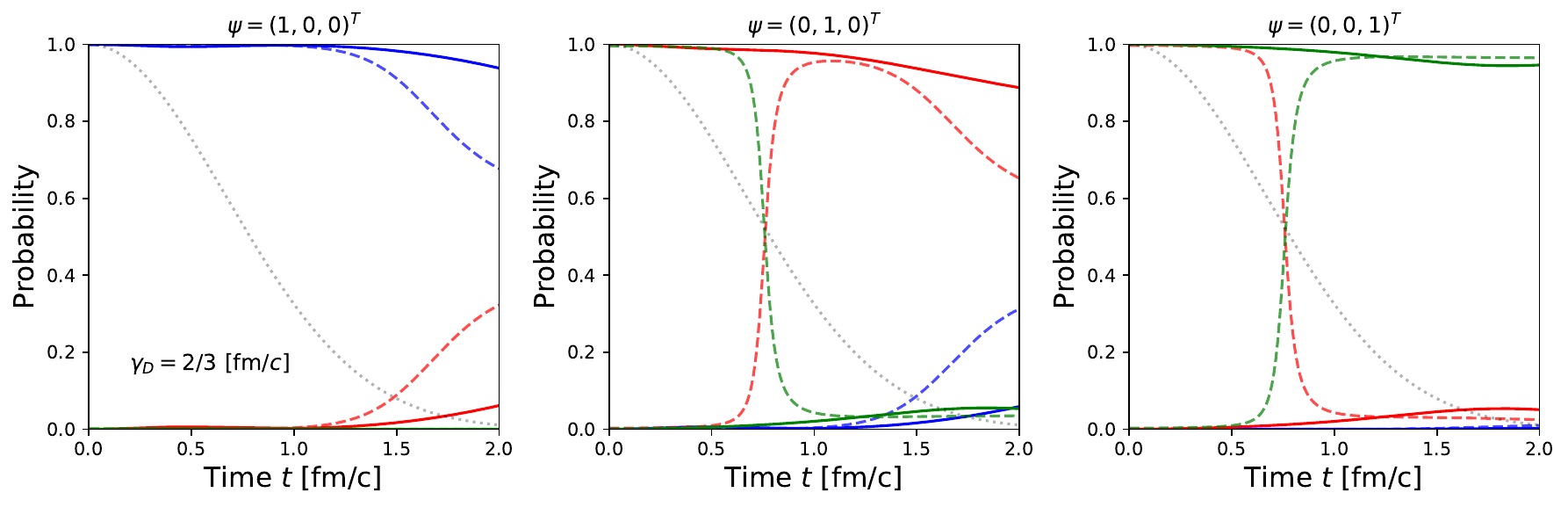}
\caption{Occupation probabilities in the three-channel model for a fast sweep, shown for three different initial diabatic states: (left) $\psi=(1,0,0)^T$ initial state, (middle) $\psi=(0,1,0)^T$ initial state, and (right) $\psi=(0,0,1)^T$ initial state.
The top and bottom panels correspond to a linear ramp and a Gaussian-decay magnetic-field profile, respectively. }
    \label{fig:single-3level-na}
\end{figure*}

Furthermore, the choice of initial state plays a crucial role in determining the occupation probabilities.
As representative examples, the system is typically initialized in a state purely occupied by one component in the diabatic basis, represented by a vector such as $\psi=(1,0)^T$ or $(0,1)^T$ in the two-channel system, corresponding to the $J/\psi$- or $\eta_c^\prime$-started states, respectively. 
In this case, the diabatic probabilities at the initial time are $P_{\rm dia}[J/\psi] = 1$, $P_{\rm dia}[\eta_c^\prime] = 0$ for the $\psi=(1,0)^T$ initial state, and vice versa for the $\psi=(0,1)^T$ initial state.
In contrast, an adiabatic (instantaneous eigenstate) initial condition is defined by taking the system to initially occupy one of the eigenstates of the Hamiltonian at the starting time. 
For example, in the two-channel system, the $\phi_{\rm adi}=(1,0)^T$ initial state is defined such that the adiabatic probabilities satisfy $P_{\rm adi}[J/\psi] = 1$ and $P_{\rm adi}[\eta_c^\prime] = 0$ at the initial time ($t=0$).
More generally, multichannel effects can significantly influence the occupation probabilities.
Including additional charmonium states leads to a coupled-channel system with multiple avoided crossings and level repulsion.
In this work, we therefore focus on the roles of the initial-state conditions and multichannel effects in determining the occupation probabilities.

\subsection{Single passage}
\label{sec:single}

We first study the dynamical evolution of charmonium states during a single passage through an avoided level crossing.
Two representative magnetic-field profiles are considered: (i) the linear ramp and (ii) the Gaussian decay.
Although their temporal behaviors differ, both profiles realize a single traversal of the avoided-crossing region and thus provide complementary realizations of LZ dynamics.
For the linear ramp, the magnetic field increases monotonically with a constant sweep rate $\nu_B$, providing the standard LZ  setup in which the adiabaticity is directly controlled by the field slope.
In contrast, the Gaussian decay features a time-dependent sweep rate, leading to an earlier traversal of the avoided crossing.

\subsubsection{Two-channel system}

To begin with, we consider the two-channel system consisting of the $J/\psi$ and $\eta_c^\prime$ states, which is an idealized situation but captures the essential features of LZ transitions.
Figure~\ref{fig:single-2level} shows the time evolution for three representative sweep rates, starting from a $\psi=(1,0)^T$ initial state.
The top panels display results for the linear ramp, while the lower panels correspond to the Gaussian-decay profile.
In all panels, the gray dotted line indicates the magnetic-field profile with a fixed maximum value of $eB_{\max}=2~\text{GeV}^2$.

\begin{figure*}[t!]
    \centering
    \includegraphics[width=0.96\textwidth]{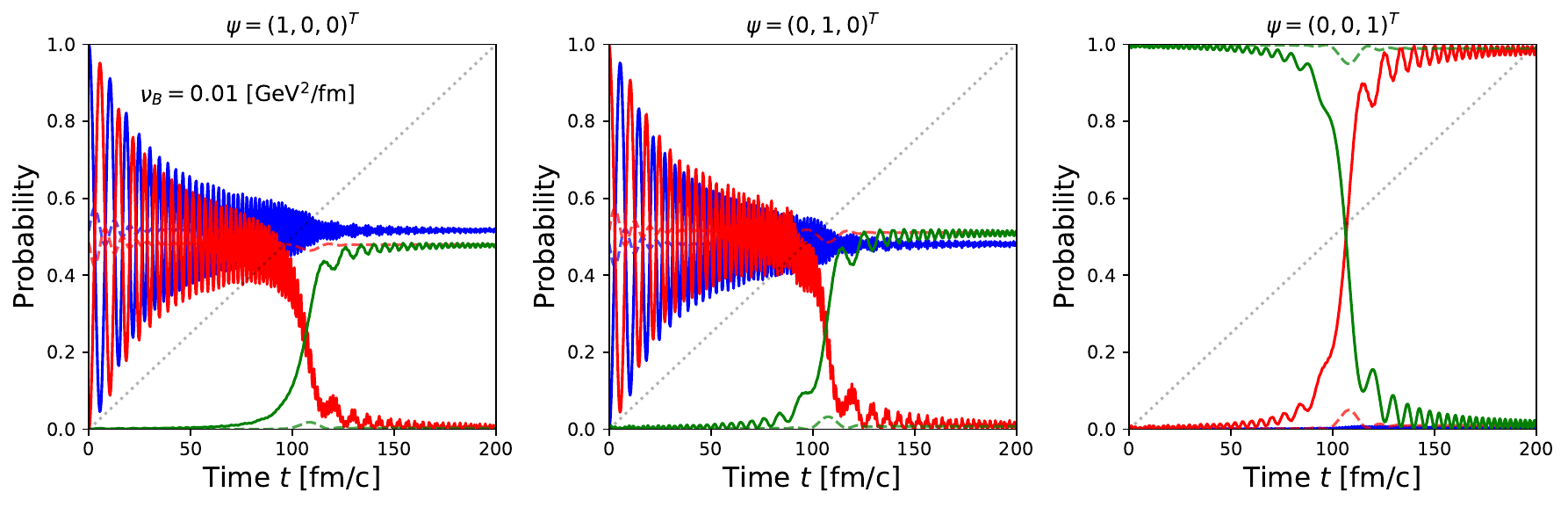}
    \includegraphics[width=0.96\textwidth]{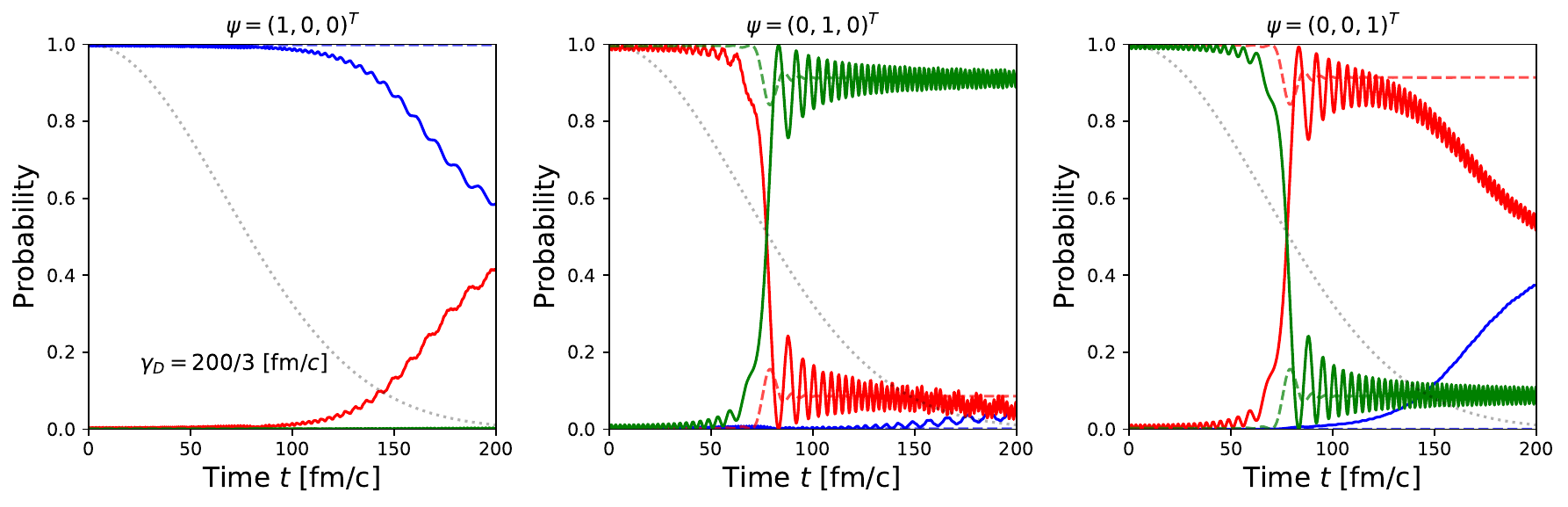}
\caption{Same as Fig.~\ref{fig:single-3level-na}, but for a slow sweep, illustrating the adiabatic evolution.}
    \label{fig:single-3level-a}
\end{figure*}

For the linear ramp (top panels of Fig.~\ref{fig:single-2level}), a fast sweep ($\nu_B=1$ GeV$^2$/fm) leads to strongly nonadiabatic evolution: the system cannot follow the instantaneous eigenstates, and the diabatic probabilities remain nearly constant, while the adiabatic probabilities undergo a sharp redistribution at the avoided crossing.
At intermediate sweep rates ($\nu_B=0.1$), nonadiabatic transitions persist, resulting in a partial reduction of $P_{\rm dia}[J/\psi]$.
For sufficiently slow sweeps ($\nu_B=0.01$), the evolution becomes adiabatic, with near-complete transfer between the diabatic channels and adiabatic probabilities that remain almost unchanged except near the crossing, in accordance with the LZ prediction of exponentially suppressed nonadiabatic transitions given in Eq.~\eqref{eq:lz-prob}.
A more detailed discussion is provided in Appendix~\ref{sec:analytic}.

The Gaussian-decay profile (bottom panels of Fig.~\ref{fig:single-2level}) exhibits similar behavior during the initial passage through the avoided crossing but differs at later times due to its time-dependent sweep rate.
Because the crossing occurs earlier than in the linear ramp, nonadiabatic effects are enhanced compared to a linear ramp with the same maximum field.
As a result, even for relatively wide Gaussian-decay profiles ($\gamma_D=200/3$ fm/$c$), the $P_{\rm dia}[J/\psi]$ does not vanish as completely as in the slow linear-ramp case.

It is worth noting that, in the evolution of diabatic probabilities $P_{\rm dia}$, small oscillations are visible in the occupation probabilities both before and after the avoided crossing.
The precrossing oscillations arise because the initial diabatic state $\psi$ is not an instantaneous eigenstate of the Hamiltonian $\phi_{\rm adi}$ and therefore decomposes into a superposition of adiabatic states with different energies.
The relative phase between these components evolves in time, producing interference patterns.
Such early oscillations are absent when the initial state is prepared in the lower adiabatic state, as discussed further in Appendix~\ref{sec:analytic}.
Moreover, the oscillation frequency is uniquely determined by the spectrum.
After the crossing, the oscillation frequency $\omega(t) = E_+(t) - E_-(t)$ increases, leading to faster but smaller-amplitude oscillations.

Taken together, the evolution of the $J/\psi$ and $\eta_c^\prime$ under a Gaussian-decaying field shares several qualitative features with the linear-ramp case.
Because the sweep rate rapidly decreases after the crossing, the system undergoes an early transition followed by a long evolution in the weak-field tail.
In contrast, the linear ramp provides only a short interference interval, resulting in weaker oscillations.
Furthermore, it is important to note again that this hadronic two-level system represents an idealized scenario that provides a controlled starting point. In more realistic situations, however, the dynamics may be affected by mixing with other nearby channels.

\subsubsection{Multichannel system}
\label{sec:multi-channel}

In the case of three or more channels, the dynamics become richer due to the presence of multiple avoided crossings and level repulsion.
For the three-channel case, the basis includes the lower $\eta_c$ state together with the $J/\psi$ and $\eta_c^\prime$ states, which is regarded as an extension of the two-channel model discussed in the previous section.
Now we can examine three different initial conditions dominated by one of the three channels.
Figure~\ref{fig:single-3level-na} shows a fast-sweep case, while Fig.~\ref{fig:single-3level-a} corresponds to a slow sweep.
In each figure, the top panels display the linear-ramp results and the bottom panels those for the Gaussian-decay profile.

\begin{figure}[t]
    \centering
    \includegraphics[width=0.96\columnwidth]{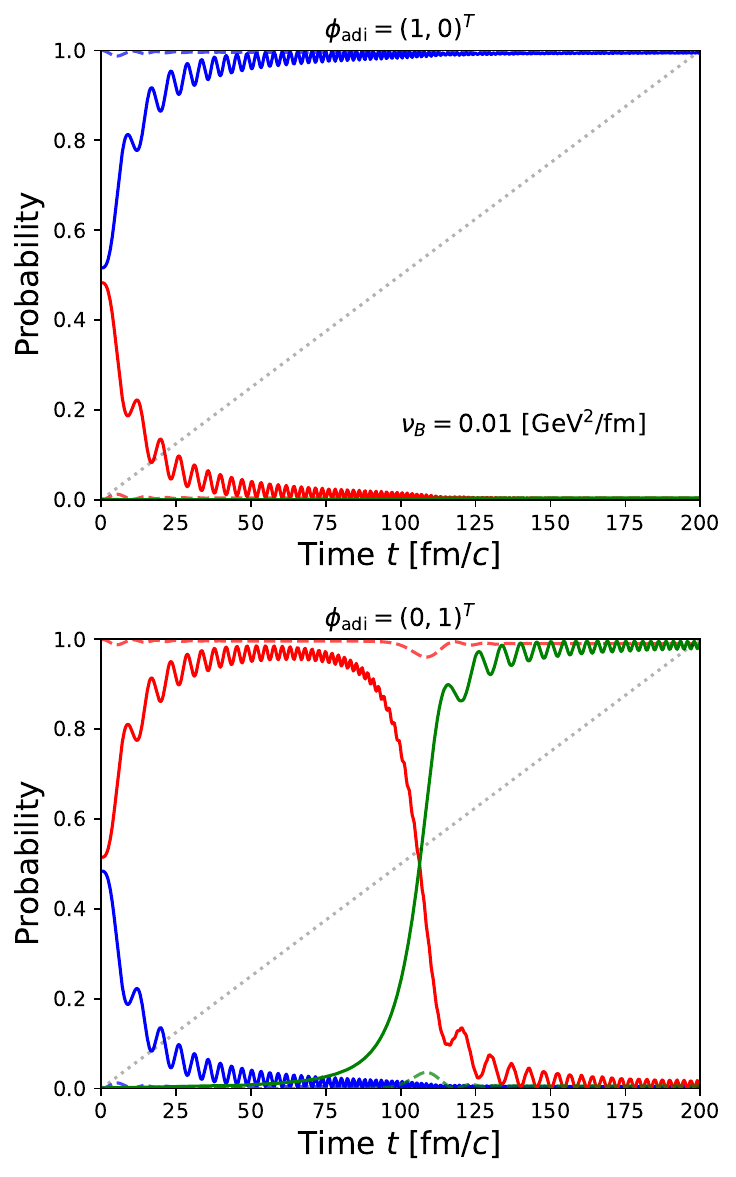}
\caption{Occupation probabilities in the three-channel model for $\phi_{\rm adi}=(1,0)^T$ and $\phi_{\rm adi}=(0,1)^T$ initial states for a slow sweep in the three-channel model. In this case, no large-amplitude oscillations appear, in contrast to those found in Fig.~\ref{fig:single-3level-a}. The results for the linear ramp now mirror those for the decaying field.}
    \label{fig:single-3level-adiabatic}
\end{figure}

In the fast-sweep regime, nonadiabatic effects dominate, and the resulting occupation dynamics exhibit a strong dependence on the initial state. For $\psi=(1,0,0)^T$ and $\psi=(0,1,0)^T$ initial states in the linear-ramp case (top panels of Fig.~\ref{fig:single-3level-na}), the adiabatic probabilities (dashed lines), $P_{\rm adi}[\eta_c]$ and $P_{\rm adi}[J/\psi]$, start near one-half, which reflect nonadiabatic transitions near the level repulsion in the weak-field region.
For the $\psi=(1,0,0)^T$ initial state, the diabatic probabilities (solid lines), $P_{\rm dia}[\eta_c]$ and $P_{\rm dia}[J/\psi]$, remain nearly unchanged, with the adiabatic probability $P_{\rm adi}[\eta_c]$ simply following $P_{\rm dia}[\eta_c]$.
For the $\psi=(0,1,0)^T$ initial state, a pronounced nonadiabatic transition occurs near the avoided crossing, where most of $P_{\rm adi}[J/\psi]$ is transferred to $P_{\rm adi}[\eta_c^\prime]$.
In contrast, for the $\psi=(0,0,1)^T$ initial state, the evolution closely resembles that of the two-channel system, with the influence of the lower $\eta_c$ level remaining weak during a single passage, reflecting the truncated nature of the three-channel model in which couplings to higher-lying states are neglected.

Comparing the magnetic-field profiles, the fast-sweep dynamics for the Gaussian-decay case shown in the bottom panels of Fig.~\ref{fig:single-3level-na} is qualitatively very similar to that of the linear-ramp case.
The main difference lies in the time structure of the evolution: for the Gaussian-decay profile, the occupation dynamics is effectively mirrored in time due to the decaying field.

\begin{figure}[t!]
    \centering
    \includegraphics[width=0.96\columnwidth]{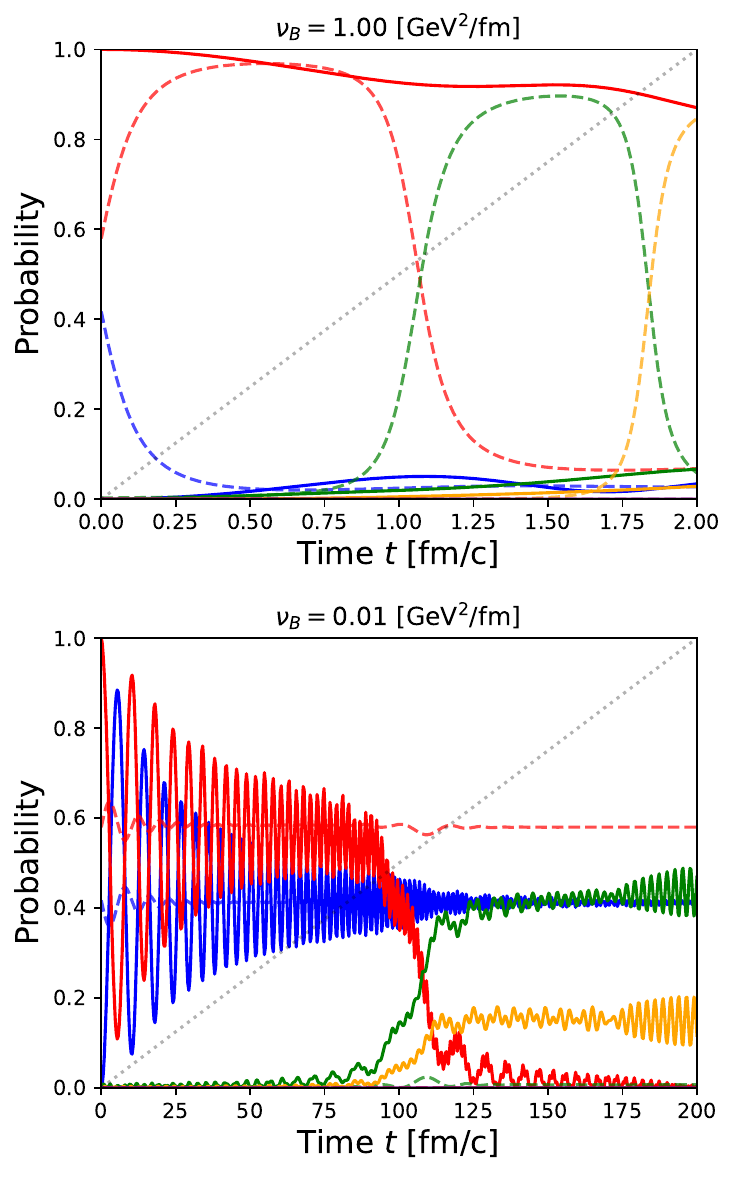}
\caption{Occupation probabilities in the five-channel model for a $\psi=(0,1,0,0,0)^T$ initial state with fast and slow linear ramps. 
 }
    \label{fig:5level-transition}
\end{figure}

For sufficiently slow sweeps, the evolution becomes predominantly adiabatic.
In this regime, the $P_{\rm dia}$ undergoes substantial redistribution, depending on the initial condition, as illustrated by the linear-ramp case shown in the top panels of Fig.~\ref{fig:single-3level-a}.
For the $\psi=(1,0,0)^T$ and $\psi=(0,1,0)^T$ initial states, early large-amplitude oscillations appear in $P_{\rm dia}[\eta_c]$ and $P_{\rm dia}[J/\psi]$.
These reflect coherence interference rather than genuine nonadiabatic transitions, since the corresponding $P_{\rm adi}$ remain essentially frozen. 
As the system evolves, $P_{\rm dia}$ is gradually transferred among the channels, leading to a large $P_{\rm dia}[\eta_c^\prime]$ after the avoided-crossing region.
For the $\psi=(0,0,1)^T$  state, the evolution closely resembles that of the two-channel system, with the lower $\eta_c$ channel playing only a minor role.

We now contrast this behavior with that obtained for the Gaussian-decay profile shown in the bottom panels of Fig.~\ref{fig:single-3level-a}.
In this case, no large-amplitude oscillations appear in the $P_{\rm dia}$.
In strong magnetic fields, the $\psi=(1,0,0)^T$ initial  state is effectively isolated from the other channels, suppressing coherent interference and leading to a smooth evolution dominated by level repulsion in the weak-field region.
For the $\psi=(0,1,0)^T$ initial state, the evolution closely resembles that of the two-channel system, whereas for the $\psi=(0,0,1)^T$ initial state, level repulsion involving the lower $\eta_c$ channel significantly affects the $P_{\rm dia}$ in the weak-field region.

In contrast, when the system is initialized in the $\phi_{\rm adi}=(1,0)^T$ and $\phi_{\rm adi}=(0,1)^T$ initial states, no large-amplitude oscillations appear in the $P_{\rm dia}$, which start instead from values around one-half, as shown in Fig.~\ref{fig:single-3level-adiabatic}. 
This difference arises because the large-amplitude oscillations observed for diabatic initial states $\psi$ (see top panels of Fig.~\ref{fig:single-3level-a}) are from coherent interference between the diabatic components.
For an adiabatic initial state $\phi_{\rm adi}$, the system initially follows the instantaneous eigenstate, leading to a relatively smooth evolution with only small oscillations.
Consequently, the resulting occupation-probability pattern exhibits a qualitatively inverted temporal structure compared with that obtained for the Gaussian-decay case.

Figure~\ref{fig:5level-transition} further illustrates these features within an extended five-channel model for a $\psi=(0,1,0,0,0)^T$ initial state.
Both fast and slow linear ramps are shown, highlighting how additional channels modify the evolution compared to the simpler three-channel case.
In particular, nonadibatic transitions involving the $\psi^\prime$ state emerges predominantly at large magnetic fields in the fast sweep.
Notably, oscillations between the $P_{\rm dia}[J/\psi]$ and $P_{\rm dia}[\eta_c]$ states persist in the slow-sweep regime even in the extended multichannel system.
Moreover, when using from the $\psi=(0,0,1,0,0)^T$ state, large-amplitude oscillations between the $\eta_c^\prime$ and $\psi^\prime$ channels also exist.

\begin{figure}[b]
    \centering
    \includegraphics[width=0.96\columnwidth]{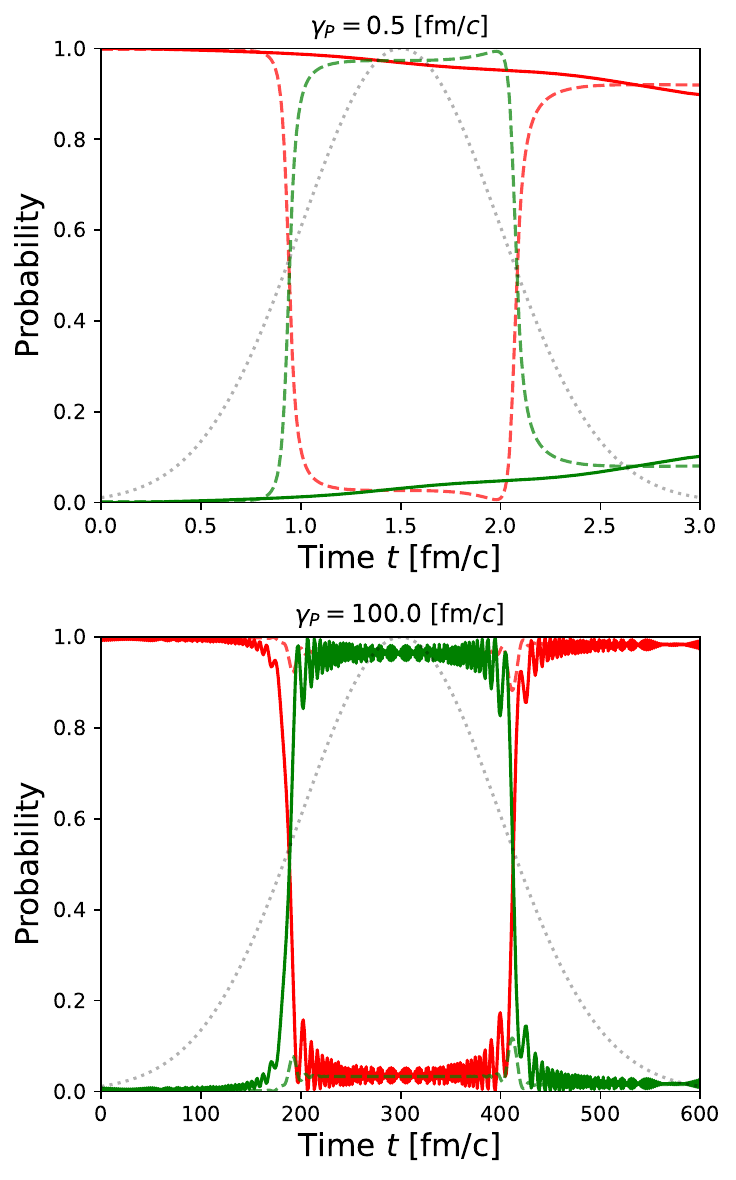}
\caption{Occupation probabilities in the two-channel model for a $\psi=(1,0)^T$ initial state with fast and slow Gaussian pulses. The gray dotted line indicates the Gaussian pulse, with the magnetic field peaking at $eB_{\max} = 2~\text{GeV}^2$.}
    \label{fig:2level-gauss}
\end{figure}

\begin{figure}[t]
    \centering
    \includegraphics[width=0.96\columnwidth]{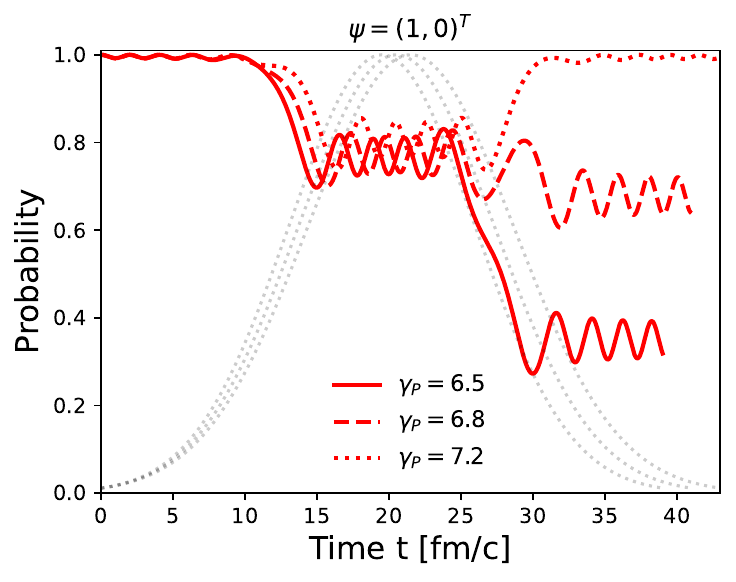}
\caption{Diabatic occupation probabilities $P_{\rm dia}[J/\psi]$ in the two-channel model for a $\psi=(1,0)^T$ initial state with three intermediate pulse widths, illustrating Stückelberg interference between the two passages and the large variation after the second passage resulting from small changes in $\gamma_P$ [fm/$c$].}
    \label{fig:gauss2-inter}
\end{figure}

Overall, the lower and higher channels play a distinct role in both the fast- and slow-sweep regimes. In the fast-sweep regime, the influence of the lower channel is minor, while the nonadiabatic transition to the higher channel become open. 
In the slow-sweep regime, however, the nearby channel becomes more significant: in the linear-ramp case, level repulsion in the weak-field region induces large-amplitude oscillations in the $P_{\rm dia}$, underscoring its role in shaping the multichannel evolution.

\begin{figure*}[t!]
    \centering
    \includegraphics[width=0.96\textwidth]{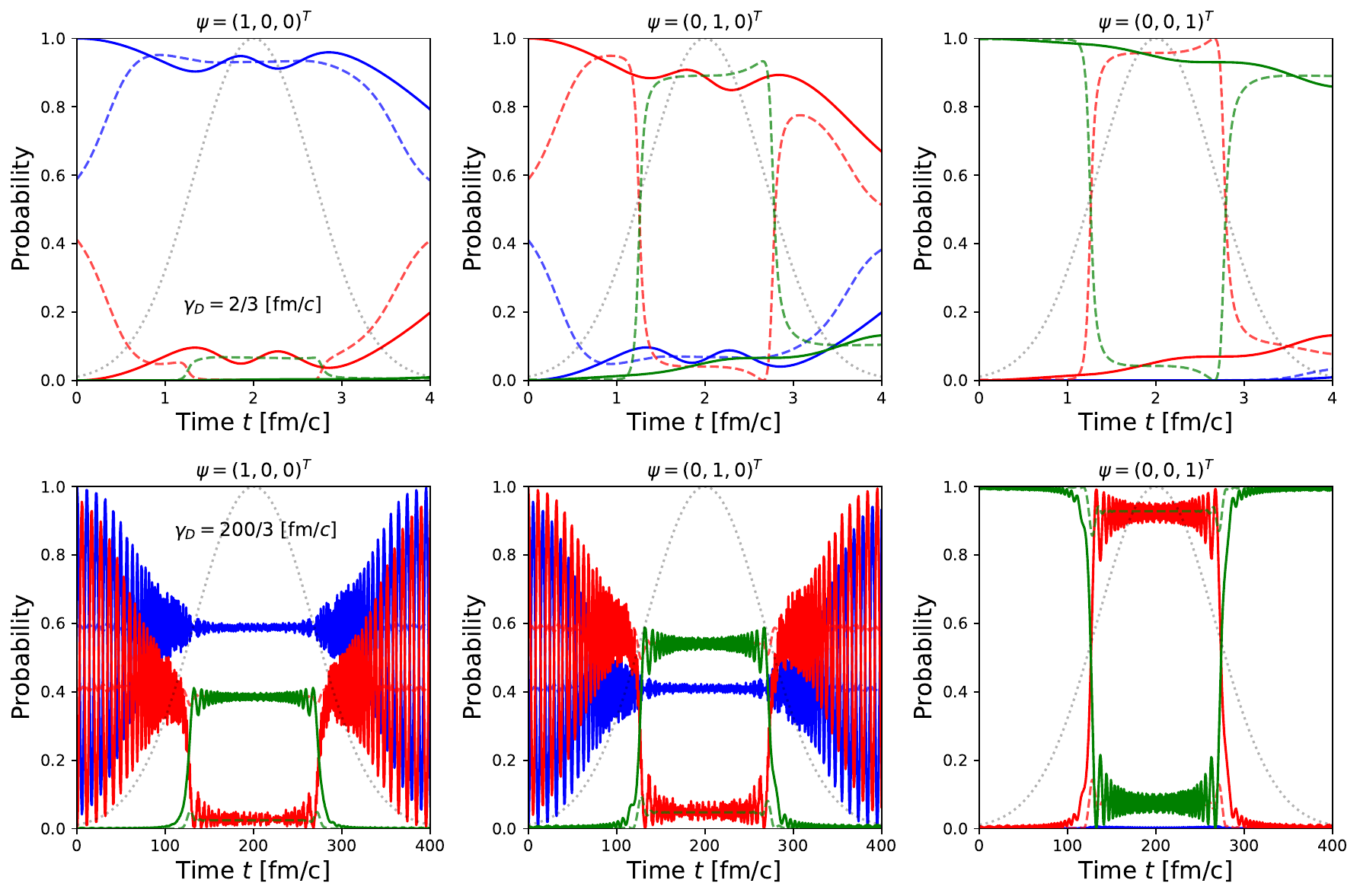}
\caption{Occupation probabilities in the three-channel model for three different initial states in a Gaussian pulse, showing fast sweep (top panels) and slow sweep (bottom panels).  
}
    \label{fig:3level-gauss}
\end{figure*}

\subsection{Double passage}
\label{sec:double}

Next, we focus on the double-passage case, using a Gaussian pulse as the magnetic-field profile. The pulse has a symmetric rise and fall around its peak, allowing the system to traverse the avoided crossings twice. This double passage can give rise to interference effects between transitions, producing more complex occupation dynamics than in the single-passage case. We also examine the influence of the nearby channel on these dynamics.

\subsubsection{Two-channel system} 

Figure~\ref{fig:2level-gauss} show the evolution for a $\psi=(1,0)^T$ initial state in the two-channel system for the fast and slow Gaussian pulses. 
For the fast pulse, the rapidly varying field induces strongly nonadiabatic dynamics, with adiabatic probabilities crossing twice.\footnote{Experimentally, such a fast Gaussian magnetic-field profile is expected to be induced by heavy-ion collisions, where produced charmonia may undergo nonadiabatic transitions.}
For the slow pulse, the first passage is largely adiabatic, leading to a strong reduction of the $P_{\rm dia}[J/\psi]$, while the second passage drives it back toward unity.

For the intermediate pulse, as depicted in Fig.~\ref{fig:gauss2-inter}, the $P_{\rm dia}[J/\psi]$ decreases during the first passage, while the probability after the second passage changes rapidly and becomes highly sensitive to the sweep rate.
For example, when we compare small variations of $\gamma_P = 6.5,~6.8,$ and $7.2$ fm/$c$, the $P_{\rm dia}[J/\psi]$ after the second passage varies sharply from 0.4 to 1.0. 
While the $P_{\rm dia}$ after the first passage can be obtained approximately using the LZ formula, the $P_{\rm dia}$ after the second passage can also be predicted via the Landau-Zener-St\"uckelberg (LZS) interference~\cite{Stuckelberg:1932} (see a discussion in great details in Appendix~\ref{sec:analytic}, where a triangular pulse is applied for simplicity).
Thus, by using the LZS interference in double passage problems, one can control the final probability of the charmonium states by tuning pulse parameters.

\subsubsection{Multichannel system}

Figure~\ref{fig:3level-gauss} shows the evolution of the three-channel system for different initial states under a Gaussian pulse. 
The left panels correspond to the $\psi=(1,0,0)^T$ initial state: for the fast pulse, $P_{\rm dia}$ changes slightly while $P_{\rm adi}$ tracks them except near the level-repulsion region; for the slow pulse, large-amplitude oscillations appear between the $P_{\rm dia}[\eta_c]$ and $P_{\rm dia}[J/\psi]$, while $P_{\rm dia}[\eta_c^\prime]$ rises at the crossing and falls afterward, with $P_{\rm adi}$ nearly constant. 
The middle panels of Fig.~\ref{fig:3level-gauss} show the $\psi=(0,1,0)^T$ initial state. 
Fast pulses induce substantial nonadiabatic transfer to $P_{\rm dia}[\eta_c^\prime]$, with minimal $P_{\rm dia}[\eta_c]$, while slow pulses follow the adiabatic path, producing probabilities similar to the $\psi=(1,0,0)^T$ case but with $P_{\rm dia}[\eta_c]$ and $P_{\rm dia}[\eta_c^\prime]$ interchanged between crossings.
For the $\psi=(0,0,1)^T$ initial state shown in the right panels of Fig.~\ref{fig:3level-gauss}, the dynamics for the slow and fast pulses resemble the two-channel $J/\psi$–$\eta_c^\prime$ case, as the lower $\eta_c$ channel has minor influence. 
Finally, if higher channels are included, the $\psi^\prime$ channel will affect the probabilities; its effects in the two-passage case can be inferred from the linear-ramp results shown in Fig.~\ref{fig:5level-transition}.

\section{Conclusion and outlook}
\label{sec:conclusion}

In this work, we have studied occupation probabilities of charmonium states in time-dependent magnetic fields by constructing multichannel Landau-Zener Hamiltonians fitted from the charmonium spectrum in static magnetic fields~\cite{Suzuki:2016kcs,Yoshida:2016xgm}.
This approach allows us to capture the essential structure of avoided level crossings and investigate the dynamical evolution of occupations under different sweep rates and magnetic-field profiles.

Our analysis demonstrates that the magnetic-field sweep rate is the key factor controlling occupation transfer among charmonium states. Rapid variations over a few fm/$c$ induce strongly nonadiabatic transitions, in accordance with the exponential Landau-Zener scaling~\cite{Majorana:1932,Landau:1932wdt,Landau:1932vnv,Zener:1932}. Intermediate sweep rates lead to partially nonadiabatic evolution with incomplete population transfer, while sweep times of several hundred fm/$c$ strongly suppress nonadiabatic transitions and result in predominantly adiabatic dynamics.

The magnetic-field profile also plays an important role. A linear ramp serves as a benchmark for single-passage dynamics, whereas a Gaussian decay requires a longer duration to reach the adiabatic regime because the avoided crossing occurs earlier. By contrast, Gaussian pulses produce a symmetric rise and fall, leading to a double passage through the avoided crossing and the emergence of Landau-Zener-St\"uckelberg interference~\cite{Stuckelberg:1932}, unlike single-passage profiles that exhibit only standard Landau-Zener transitions~\cite{Majorana:1932,Landau:1932wdt,Landau:1932vnv,Zener:1932}.

Multichannel effects and the choice of initial state further influence the dynamics: including the nearby $\eta_c$ and $\psi^\prime$ states opens additional transition pathways, producing richer interference patterns, while coherent interference in the initial state generates large-amplitude oscillations, particularly in the adiabatic regime.
Such behavior appears to be characteristic of multichannel dynamics in hadronic systems.

\begin{figure*}[t!]
    \centering
    \includegraphics[width=0.96\textwidth]{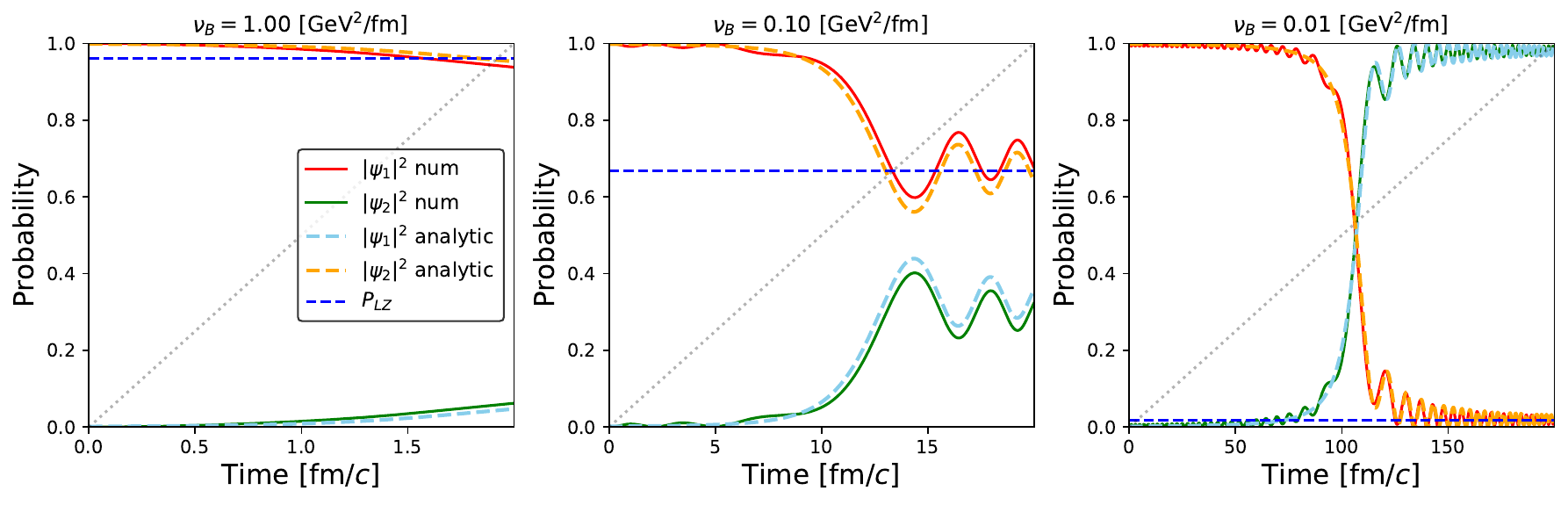}
\caption{Comparison between the analytic Landau-Zener prediction and the numerical solution of the time-dependent Schr\"odinger equation for several different sweep rates. The analytic result assumes that the system is prepared in the lower adiabatic initial state, not in the diabatic initial state. This difference in initial conditions leads to small discrepancies, most notably the absence of precrossing oscillations in the analytic curves. The horizontal blue dashed line marks the analytic Landau-Zener prediction, which agrees well with the numerical evolution for each sweep rate.}

    \label{fig:2analytic}
\end{figure*}

We emphasize that a key advantage of the Landau-Zener framework is that it does not require specifying a detailed microscopic Hamiltonian; instead, it adopts a spectrum-driven perspective. In this work, we have used the quark-model spectrum as an input. Nevertheless, the spectrum is not tied to any specific model and can be obtained from other approaches, such as lattice QCD or effective field theories.
In this sense, the LZ framework is a versatile tool for studying nonadiabatic transitions in quarkonium systems.

This study demonstrates the usefulness of Landau-Zener Hamiltonians for understanding and predicting nonadiabatic charmonium dynamics in strong and time-dependent magnetic fields.
These results can serve as a guideline for future lattice QCD simulations of charmonia under time-dependent magnetic fields.
We note that real-time evolution in Monte-Carlo simulations is notoriously difficult due to the sign problem.
However, using sign-problem-free approaches, the real-time dynamics of one-dimensional systems containing both gauge and fermionic degrees of freedom has been extensively investigated.
For example, in the Schwinger model, a (1+1)-dimensional U(1) gauge field coupled to dynamical fermions, the real-time dynamics of physical quantities has been extensively simulated using tensor-network methods, e.g.,~\cite{Pichler:2015yqa,Rigobello:2021fxw,Belyansky:2023rgh}.
The physical phenomena induced by time-dependent magnetic fields studied in this work
are therefore expected to serve as representative benchmarks for simulations in two
or higher spatial dimensions.

It is also important to distinguish nonadiabatic transitions from decays into different asymptotic states.
Nonadiabatic transitions arise from coherent magnetic-field–induced mixing between the two charmonium bound states and merely redistribute probability amplitudes without changing the particle content or total probability.
In contrast, decays into other asymptotic states are irreversible processes that change the number and type of particles and lead to probability loss from the two-level subsystem.
In the present analysis, we neglected realistic decay effects and leave a more complete treatment including finite lifetimes for future work.

\section*{Acknowledgment}

A.J.A. was supported by the JAEA Postdoctoral Fellowship Program, and partially by the RCNP Collaboration Research Network Program under Project No. COREnet 057, as well as by the PUTI Q1 Grant from the University of Indonesia under Contract No. PKS-206/UN2.RST/HKP.05.00/2025.
This work was supported by the Japan Society for the Promotion of Science (JSPS) KAKENHI (Grant No. JP24K07034).

\appendix
\section{Analytic solutions}
\label{sec:analytic}

In this Appendix, we discuss the relationship between the analytic solutions known in the LZ problem and our numerical results shown in the main text.

\subsection{Single passage}

For the numerical analysis, we consider a two-level Hamiltonian of the form  
\begin{equation}
H(t)=
\begin{pmatrix}
\alpha_1 eB(t) + \delta_1 & \Delta \\
\Delta & \alpha_2 eB(t) + \delta_2
\end{pmatrix},  
\quad 
eB(t)=\nu_B t ,
\end{equation}
which represents a linear magnetic-field sweep with an avoided level crossing.  
The instantaneous level spacing (at $\Delta=0$) is
$\nu_{\rm eff}=|\alpha_1-\alpha_2|\nu_B$ .
The crossing occurs at the shifted time
\begin{equation}
t_c=\frac{\delta_2-\delta_1}{ (\alpha_1-\alpha_2)\nu_B }.
\end{equation}

To compare with the numerical solution of the time-dependent 
Schr\"odinger equation, we use the exact LZ wave functions written in
terms of parabolic cylinder functions.  
Defining the dimensionless variables
\begin{equation}
\tau = \sqrt{\nu_{\rm eff}}\,(t - t_c), 
\qquad 
\lambda=\frac{\Delta^2}{|\nu_{\rm eff}|},
\end{equation}
the analytic amplitudes are
\begin{align}
\psi_2(t) &= 
D_{-i\lambda}\!\left( e^{-3\pi i/4}\tau \right),\\
\psi_1(t) &=
\frac{\Delta}{\sqrt{\nu_{\rm eff}}}\,
D_{-1-i\lambda}\!\left( e^{-3\pi i/4}\tau \right),
\end{align}
where $D_\nu(z)$ is the parabolic cylinder function.
The overall normalization is fixed at each time by
$|\psi_1|^2 + |\psi_2|^2 = 1$,
ensuring a direct comparison with the numerical evolution.

\begin{figure*}[t!]
    \centering
    \includegraphics[width=0.96\textwidth]{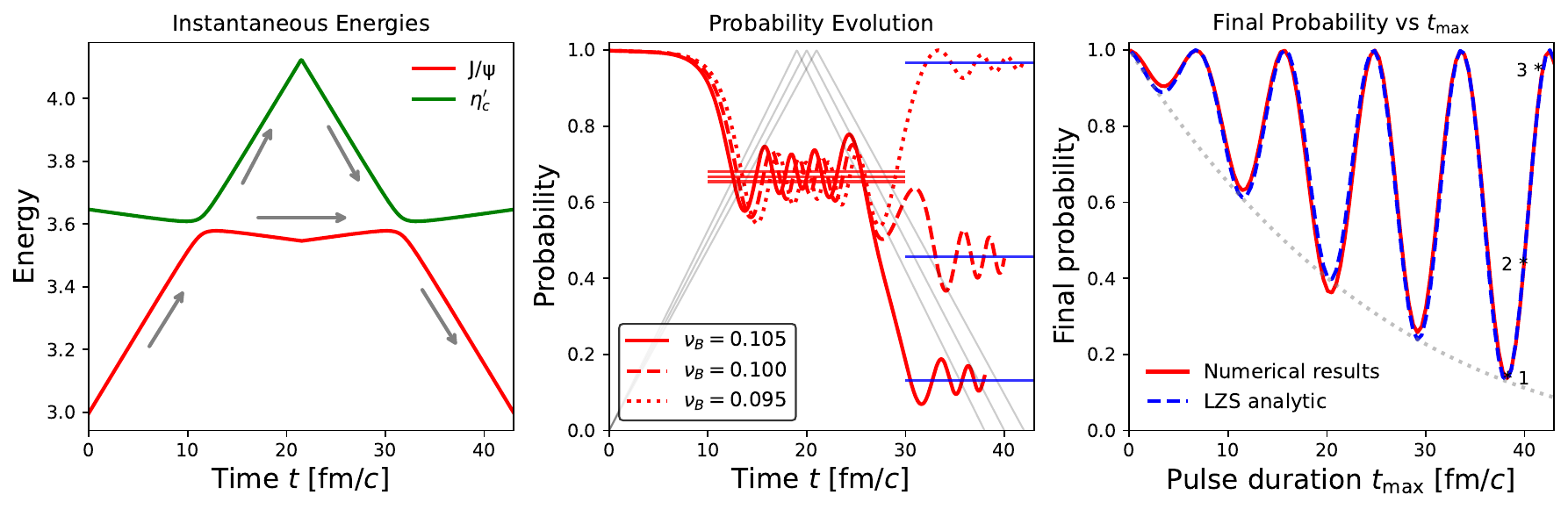}
\caption{
Double-passage LZS interference for a
triangular magnetic-field pulse.
Left panel: instantaneous adiabatic energies $E_\pm(t)$ showing two
avoided crossings.
Middle panel: time evolution of the diabatic probabilities for
three pulse durations, illustrating Stückelberg oscillations generated
by the interference between the amplitudes from the two passages.
Right panel: final transition probability as a function of the pulse
duration $t_{\max}$.
The numerical results are compared with the analytic LZS
prediction. Some points are indicated for the process plotted in the middle panel.
}

    \label{fig:lzs-double}
\end{figure*}

The analytic solution assumes that the system is prepared in the lower adiabatic state.
In contrast, our numerical evolution starts 
with the lower \emph{diabatic} state, $\psi(0)=(1,0)^T$, which is not an instantaneous eigenstate of the Hamiltonian.
Therefore, this state is represented as a superposition of two states in the adiabatic basis.
Since the two adiabatic components have different energies, the relative phase contained in the interference term evolves in time as $\Delta\phi(t)=\int_0^t [E_+(t')-E_-(t')]dt'$, leading to oscillations in the occupation probabilities before the avoided crossing.
After the crossing, the interference between amplitudes that followed different adiabatic paths can also produce oscillations. 
However, the finite-time initialization
causes the oscillation values in the numerical solution to deviate slightly from the ideal analytic result, leading to the minor discrepancies observed in Fig.~\ref{fig:2analytic}.

Matching the asymptotic forms of the parabolic cylinder functions
gives the standard LZ probability~\cite{Majorana:1932,Landau:1932wdt,Landau:1932vnv,Zener:1932,Stuckelberg:1932}
\begin{equation}\label{eq:lz-prob}
P_{\mathrm{LZ}}
= \exp\!\left(-2\pi\lambda\right),
\end{equation}
which we use as a benchmark for our numerical single-passage results.
It is important to note, however, that this analytic expression strictly applies only when the system is prepared in a pure diabatic initial state and the dynamics involve a single isolated avoided crossing. This expected consistency is confirmed in Fig.~\ref{fig:2analytic}, where the numerical results approaches the analytic LZ formula.

\subsection{Double passages}

When the magnetic field drives the system through the avoided crossing twice, as in the triangular pulse considered here for analytic discussion, the evolution consists of two successive LZ transitions. Each passage coherently splits the wave function between the adiabatic branches (left panel of Fig.~\ref{fig:lzs-double}), creating two pathways. These pathways accumulate different phases along their respective adiabatic trajectories, and their recombination at the second crossing leads to constructive or destructive interference depending on the total phase. This coherent two-path interference is the origin of the LZS oscillations~\cite{Stuckelberg:1932}.

For a triangular pulse, each level crossing determines the corresponding LZ parameter ($i=1,2$),
\begin{equation}
\lambda_i = \frac{\Delta^{2}}{|\nu_{\rm eff}^{(i)}|},
\qquad
P_{\rm LZ}^{(i)} = e^{-2\pi\lambda_i},
\end{equation}
where the effective sweep rate is  
\begin{equation}
\nu_{\rm eff}^{(i)} = |\alpha_1-\alpha_2|\;\nu_B^{(i)}.
\end{equation}
The rising and falling segments have opposite slopes, but
$P_{\rm LZ}^{(1)} = P_{\rm LZ}^{(2)}$.  
For a given total duration $t_{\max}$, the sweep rate for each single passage becomes  
\begin{equation}
    \nu_B^{(1)}=-\nu_B^{(2)} = \frac{2B_{\max}}{t_{\max}}.
\end{equation}
The middle panel of Fig.~\ref{fig:lzs-double} shows the probability evolution of the $J/\psi$ diabatic state. After the first passage through the avoided crossing, the probability is consistent with the LZ formula $P_{\rm LZ}$ (red horizontal lines) for each sweep rate $\nu_B$.
After the second passage, however, the probability begins to oscillate as a function of $\nu_B$, which is characterized by the LZS interference between the two transition pathways.

For a general two-passage process with possibly different sweep rates, the transition probability is  
\begin{align}
P_{\mathrm{LZS}} =&\;
4\sqrt{P_{\rm LZ}^{(1)}\!\left(1-P_{\rm LZ}^{(1)}\right)
      P_{\rm LZ}^{(2)}\!\left(1-P_{\rm LZ}^{(2)}\right)} \nonumber\\[3pt]
&\times 
\sin^{2}\!\left[\frac{\phi_{\rm dyn}
+\phi_{{\rm S},1}
+\phi_{{\rm S},2}}{2}\right],
\label{eq:LZS-general}
\end{align}
which reduces to the familiar expression  
\begin{equation}
    P_{\rm LZS}
=4P_{\rm LZ}(1-P_{\rm LZ})
\sin^{2}\!\left[\frac{\phi_{\rm dyn}+2\phi_{\rm S}}{2}\right]
\end{equation}
when the two passages are identical.
The two crossings occur at 
\begin{align}
t_1 =& t_c=\frac{\delta_2-\delta_1}{ (\alpha_1-\alpha_2)\nu_B }, \\
t_2 =& t_{\max}-t_c, 
\end{align}
and the accumulated dynamical phase between them is  
\begin{equation}
\phi_{\rm dyn}
= \int_{t_1}^{t_2}
\!\bigl[E_+(t)-E_-(t)\bigr]\,dt.
\end{equation}
Each individual LZ transition also produce the so-called Stokes phase~\cite{Nakamura:1984,Kayanuma:1993},
\begin{equation}
\phi_{{\rm S},i}
=
\frac{\pi}{4}
+ \lambda_i(\ln\lambda_i - 1)
+ \arg \left[\Gamma(1-i\lambda_i)\right],
\end{equation}
where $\phi_{{\rm S},i}\to\frac{\pi}{4}$ in the nonadiabatic limit ($\lambda_i\to0$), and $\phi_{{\rm S},i}\to 0$ in the adiabatic limit ($\lambda_i\to \infty$).
The right panel of Fig.~\ref{fig:lzs-double} shows the characteristic LZS oscillations obtained from our numerical simulation, compared with the analytic prediction (plotted as $1-P_{\mathrm{LZS}}$).  
The gray dotted line corresponds to the probability obtained by omitting the sine term, i.e.\ the part determined solely by the single-passage probabilities $P_{\rm LZ}$.  
The oscillatory behavior therefore originates from the accumulated phase between the two crossings.

\bibliography{references}

\end{document}